\documentclass{article}

\usepackage{arxiv}
\usepackage[utf8]{inputenc}   % Allow utf-8 input
\usepackage[T1]{fontenc}      % Use 8-bit T1 fonts
\usepackage{mathptmx}         % Times New Roman-like font (standard for many journals)
\usepackage{microtype}        % Improves character spacing/justification

\usepackage{amsfonts}         % Blackboard math symbols
\usepackage{bm}               % Bold math symbols
\usepackage{nicefrac}         % Compact symbols for fractions

\usepackage{booktabs}         % Professional-quality tables
\usepackage{threeparttable}   % Better table notes/formatting
\usepackage{graphicx}         % Include figure files
\usepackage{dcolumn}          % Align table columns on decimal point

\usepackage{etoolbox}         % Programming tools for LaTeX hooks
\usepackage{xcolor}           % Color support
\usepackage[normalem]{ulem}   % Underlining and strikethroughs
\usepackage{url}              % Simple URL typesetting
\usepackage{hyperref}         % Hyperlinks
\usepackage{float}
\usepackage{amsmath}
\usepackage{siunitx}
\usepackage{fancyhdr}
\makeatother

\title{High-intensity attosecond beamline for XUV pump – XUV probe measurements with photon energies up to \unit{150.{e}V}}

\author{
    Sajjad Vardast \\
    Department of Physics, Ume\r{a} University \\
    Ume\r{a}, 901 87, Sweden \\
    \and
    Alexander Muschet \\
    Department of Physics, Ume\r{a} University \\
    Ume\r{a}, 901 87, Sweden \\
    \and
    N. Smijesh \\
    Department of Physics, Ume\r{a} University, Sweden \\
    Raman Research Institute, Bangalore, India \\
    \and
    Mohammad Rezaei-Pandari \\
    Department of Physics, Ume\r{a} University \\
    Ume\r{a}, 901 87, Sweden \\
    \and
    Fritz Schnur \\
    Department of Physics, Ume\r{a} University \\
    Ume\r{a}, 901 87, Sweden \\
    \and
    Robin Weissenbilder \\
    Department of Physics, Lund University \\
    Lund, 221 00, Sweden \\
    \and
    Elisa Appi \\
    Department of Physics, Lund University \\
    Lund, 221 00, Sweden \\
    \and
    Jan Lahl \\
    Department of Physics, Lund University \\
    Lund, 221 00, Sweden \\
    \and
    Sylvain Maclot \\
    Department of Physics, Lund University, Sweden \\
    CIMAP UMR6252, Normandie Univ, France \\
    \and
    Per Eng-Johnsson \\
    Department of Physics, Lund University \\
    Lund, 221 00, Sweden \\
    \and
    Anne L'Huillier \\
    Department of Physics, Lund University \\
    Lund, 221 00, Sweden \\
    \and
    Laszlo Veisz \\
    Department of Physics, Ume\r{a} University \\
    Ume\r{a}, 901 87, Sweden \\
    \texttt{laszlo.veisz@umu.se}
}

\begin{document}
\maketitle

\begin{abstract}
The field of attosecond physics has expanded significantly in recent years, yet experimental facilities supporting attosecond pump -- attosecond probe spectroscopy remain rare. Here, we present a newly constructed beamline for the generation and application of energetic, isolated extreme ultraviolet (XUV) and soft X-ray attosecond pulses via upscaling of high-harmonic generation (HHG) in a gas medium. The fundamental properties of the HHG radiation ---energy, beam profile, spectrum, and divergence--- are characterized and optimized. The source delivers up to 55 nJ of pulse energy within the Zr window (65-150 eV) with high stability ($\sim$5-10\%) and a divergence of 0.1~mrad. Numerical simulations identify optimal operating conditions consistent with experimental results. Temporal super-resolution of the driving laser is applied, resulting in a broadened spectral continuum. Furthermore, the beamline includes a split-and-delay stage before focusing the HHG radiation to a $<\!6$ µm spot for pump-probe experiments using two distinct focusing optics. Spatially resolved ion microscopy is employed to trace the generated ions at the focus. \textcolor{black}{The presented beamline is designed for nonlinear XUV studies with attosecond isolated pulses.}

\end{abstract}

\flushbottom
\maketitle

\thispagestyle{empty}

\section{Introduction}

%The motion of electrons plays a fundamental role in the modern society. It is not only what drives our electronic devices, but it is also essential e.g. for the initiation of chemical reactions \cite{krausz2009attosecond, doi:10.1126/science.242.4886.1645}. Hence, it is of crucial importance to study and understand it to overcome fundamental limitations of current technology. However, the electron motion that drives present day electronic detectors is much faster than those detectors themselves. Therefore, the only way to observe these processes is via the use of very short light pulses on the attosecond regime \cite{krausz2009attosecond, kienberger2004atomic}.
%For many measurements with ultrashort light pulses, lasers are nowadays used \cite{doi:10.1126/science.242.4886.1645, stolow2004femtosecond, xu1996route}. However, the short wavelengths that are necessary to reach attosecond time resolution cannot be directly generated by a laser. Therefore, the spectrum of the laser has to be significantly broadened 

\begin{figure}
\includegraphics[width=0.7\linewidth]{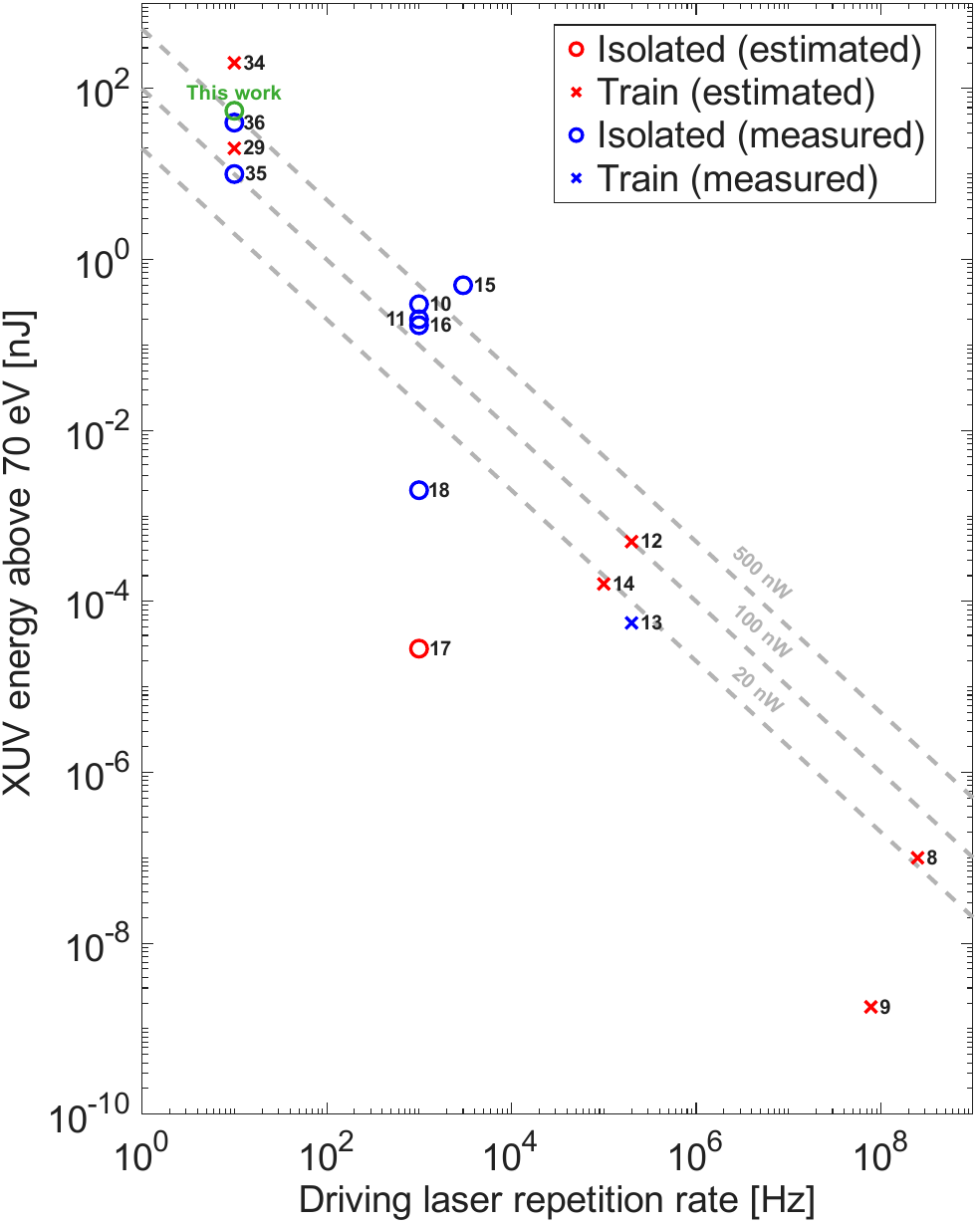}% Here is how to import EPS art
\caption{\textcolor{black}{XUV energy per pulse or train (multiple pulses) for high-order harmonic generation sources that use neon as a medium for HHG with a spectrum over 70 eV. The driving laser's repetition rate varies from 10 Hz to 250 MHz, with a central wavelength in the 800-1040~nm range, except for two cases having around 1800~nm \cite{Cousin2017,Johnson2018}. Sources marked with 'o' have reported isolated XUV pulses, while those marked with 'x' have trains of XUV pulses. The red sources indicate an estimation of XUV energy per shot, whereas the blue sources measured the XUV energy directly. The highest pulse energy is demonstrated in this work (taking into account the large number of pulses in ref~[34].}}
\label{fig:HHGR} 
\end{figure}

Utilizing high-harmonic generation (HHG) in a gas medium\cite{ferray1988multiple, mcpherson1987studies}, extreme ultraviolet (XUV) and soft X-ray pulses with a pulse duration well below the 100 as timescale are produced\cite{krausz2009attosecond, gaumnitz2017streaking, li2020attosecond,Lamas2025}. These pulses are applied in pump-probe techniques for the observation of ultrafast electron dynamics in matter, where one ultrashort light pulse starts a process, e.g. by ionizing an atom or molecule, and a second pulse probes it with a variable time delay. Therefore, at least two photons are absorbed (one from each pulse) \cite{Orfanos2019}.
However, most HHG-based attosecond sources create very low pulse energies, originating from the low conversion efficiency of the generation process\cite{krausz2009attosecond}. \textcolor{black}{This is illustrated for neon gas and XUV photon energies above 70~eV in Fig.~\ref{fig:HHGR}, where enhancement cavity-based XUV sources provide energies in the sub-fJ range with repetition rates reaching hundreds of MHz \cite{Carstens2016,Pupeza2013}. On the other hand, 1-200 kHz lasers with a central wavelength in the NIR spectral region deliver energies ranging from sub-pJ to nJ \cite{Li2019,Timmers2016,Mikaelsson2021,Harth2018,Chevreuil2021,Goulielmakis2008,Mashiko2008}. Additionally, 1 kHz lasers operating at a central wavelength around 1800 nm yield XUV energies that vary from sub-pJ to a few pJ level, though extending to higher photon energies \cite{Cousin2017,Johnson2018}. These sources} do not reach the necessary intensity for the nonlinear interaction with the target, which is essential for the pump-probe technique\cite{sansone2011high}. Hence, the majority of attosecond measurements use the fundamental laser as pump or as probe and the attosecond XUV/soft X-ray pulse as the other\cite{schultze2010,ferancesca2014,Leshchenko2023}. This does not compromise the temporal resolution because the attosecond pulse is much shorter than half an optical laser cycle, and the electric field of the laser, that changes on the attosecond timescale, governs the process and not its intensity envelope. Generating a strong signal often requires a high laser electric field on the sample, which may alter or even damage the material in contrary to intense XUV pulses in XUV pump--XUV probe measurements \cite{Palacios2009, Palacios2014}. 

%Fig.~\ref{fig:spectemp}%
%%%%%%%%%%%%%%%%%%%%%%%%%%%%%%%%%%%%%%%%%%%%%%%%%

To measure the motion of electrons without the presence of a laser field, intense XUV and X-ray pulses with attosecond time resolution are needed. One novel way to accomplish this is the use of free-electron lasers, but due to their narrow relative bandwidth of typically less than 1\% of the central photon energy, these sources only reach attosecond pulse durations mostly in the X-ray spectral region\cite{duris2020tunable}. Furthermore, the construction and operation of these free-electron lasers is extremely expensive, which is why there are only a few of these sources worldwide\cite{feng2018review}, and the first experimental studies are currently being conducted\cite{Driver2024}.

%%%%%%%%%%%% FIG 2 %%%%%%%%%%%%%%%%%%%%%%%%%
%Fig.~\ref{fig:spectemp}%
\begin{figure*}[!t]
\includegraphics[width=\linewidth]{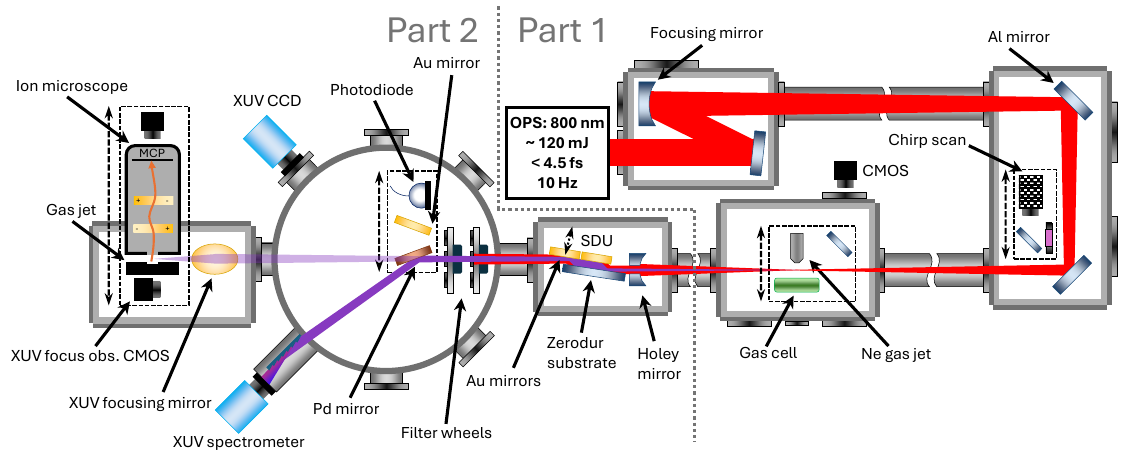}% Here is how to import EPS art
\caption{Schematic of AS beamline in the REAL lab. Part of the LWS100 laser system is guided into the beamline. It is focused by a focal length of 22~m in the first vacuum chamber. Then it is turned backwards in the next vacuum chamber, where hardware for temporal characterization of the laser pulses is placed. In the third chamber, HHG is generated in neon gas. After propagation \textcolor{black}{through the holey mirror and split-and-delay unit (SDU)}, the laser is filtered out by a substrate and metallic filters, and the basic characterization of the XUV pulse is carried out in the \textcolor{black}{fifth} chamber. An ellipsoidal gold-coated XUV focusing mirror and an ion microscope are positioned at the end of this chamber. OPS: Optical Parametric Synthesizer, CMOS: Complementary Metal-Oxide-Semiconductor, CCD: Charge-Coupled Device, MCP: Microchannel Plate. Part 1: XUV generation (see Chapter II), Part 2: XUV characterization (Chapter III).}
\label{fig:scheme} 
\end{figure*}
%%%%%%%%%%%%%%%%%%%%%%%%%%%%%%%%%%%%%%%%%%%%%%%%%

Alternatively, intense XUV and soft X-ray pulses with pulse durations on the attosecond timescale can be produced on a laboratory scale via the use of energy upscaling of HHG \cite{heyl2016scale}. Therefore, significant effort is taken to enable this alternative source 
for XUV pump--XUV probe experiments \cite{Rudawski2013, Takahashi2013, Fabris2015, bergues2018tabletop, makos2020alpha}. For this method, \textcolor{black}{sufficient XUV energies exceeding 10~nJ are needed, which can only be produced by lasers with high pulse energy ($\gg 1$~mJ) and typically low repetition rate \cite{Takahashi2004,Fu2020,Rivas2018,Rudawski2013} (Fig.~\ref{fig:HHGR})}. Furthermore, for the efficient generation of isolated attosecond pulses, very short laser pulses with less than two optical cycles are necessary. \textcolor{black}{Fulfilling} these parameters is very challenging because most high-intensity lasers, which are often based on the Ti:Sapphire amplifier medium, do not provide the pulse duration for the direct generation of isolated attosecond pulses via HHG \cite{wall1990titanium}. These systems need sophisticated techniques that significantly reduce the HHG efficiency \cite{mashiko2008double}. Alternatively, post compression methods, which are widely used for the generation of isolated attosecond pulses, are applied, but they are limited to pulse energies in the few mJ range \cite{nagy2021high}. This limits the energy of the attosecond pulses that can be produced with such sources.

Novel amplification schemes like optical parametric chirped pulse amplification (OPCPA)\cite{Kretschmar2020,Herrmann2009} and optical parametric synthesis (OPS)\cite{lws100} are fulfilling the requirements for both pulse duration and energy. OPCPA allows the amplification of a broad spectrum. Furthermore, the amplified spectral region can be adjusted over a wide range in the visible and infrared regions by tailoring the amplification conditions. Therefore, more and more OPCPA systems are being constructed right now for the purpose of generating intense XUV pulses \cite{kuhn2017eli, midorikawa2022progress,Squibb2026}. However, for the generation of well isolated attosecond pulses without the use of additional gating techniques the pulse duration that can be reached via OPCPA is still not sufficient. OPS allows the amplification of even broader spectra via the combination of several OPCPA stages that amplify different spectral regions. This finally allows the generation of sub-two-cycle laser pulses with energies of hundreds of mJ, which provides the perfect conditions for the generation of intense isolated attosecond pulses \cite{rivas2017next, Rossi2020, Xu2024, lws100}.

In this article, we present an XUV and soft X-ray light source together with its dedicated beamline, capable of generating attosecond pulses with energies of several tens of nJs and photon energies up to 150 eV. These XUV pulses are well isolated at the highest photon energies. In addition to comprehensive spectral and spatial characterization of the generated pulses, the presented beamline features a split-and-delay unit~\textcolor{black}{\cite{Campi2016}}, multiple focusing configurations, and an ion microscope. The performances of the beamline are confirmed by numerical HHG simulations. Under these conditions, intensities sufficient for nonlinear ionization of xenon are achieved \cite{bergues2017towards, muschet2021non}, making this source an ideal platform for XUV pump--XUV probe investigations.

\section{XUV generation}

To generate isolated attosecond pulses with high pulse energy via HHG in a gas medium, the driving laser is of crucial importance. For this XUV source, the Light Wave Synthesizer 100 (LWS100) system at the Relativistic Attosecond Physics Laboratory (REAL) at Umeå University was used \cite{lws100}. This laser system utilizes two color-pumped OPS to amplify laser pulses with an energy of up to 480 mJ and a spectral range from 580 nm to 1030 nm. This enables the generation of $<\!4.5$~fs light pulses with 100~TW peak power. To prevent excessive ionization, only part of the full laser energy (up to 120 mJ) is guided into the HHG beamline by clipping the beam before the compressor chamber with an iris. The reduced beam diameter is slightly tunable around 50~mm from the original 75~mm, i.e., a decrease of on-target energy to about 45\% of the original laser energy with full diameter, which is optimal for HHG. This energy is at least two orders of magnitude higher than the energy of the systems that are typically used for HHG \cite{Osolodkov2020,Lorek20214,Rothhardt2016}. Due to the short duration of those visible and near-infrared (VIS-NIR) light pulses, only 1.6 optical cycles of the electric field fit within its \textcolor{black}{full-width-at-half-maximum} (FWHM) pulse duration. This supports the generation of isolated attosecond pulses via HHG without the use of additional gating techniques that typically lower the conversion efficiency \cite{sola2006controlling, mashiko2008double,abel2009isolated}.

\begin{figure}
\centering
\includegraphics[width=\linewidth]{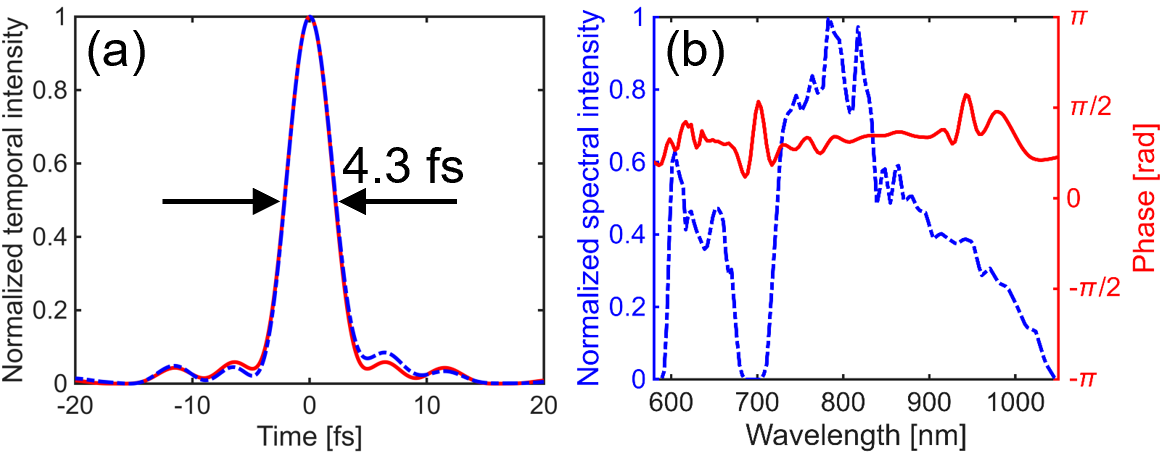}% Here is how to import EPS art
\caption{\label{fig:time} Temporal and spectral characterization of the fundamental laser pulses using the chirp-scan technique. (a) Temporal structure of the Fourier transform limited pulse (solid red) and the measured pulse (dashed blue). Both have a FWHM duration of 4.3~fs. (b) Spectrum (dashed blue) and spectral phase (solid red).}
\end{figure}

The schematic of the attosecond (AS) beamline is shown in Fig. \ref{fig:scheme}. The beamline is divided into two main sections: I) XUV generation, II) XUV characterization and application. The XUV generation section includes focusing optics and characterization of the driving laser, as well as a gas source for HHG, while the second section includes XUV characterization elements, XUV focusing optics, and an ion microscope. The XUV generation section has a vacuum system that is matched to the 22 m focusing geometry, including a chamber with focusing optics, two approximately 11-meter beam path tubes, and a turning chamber that redirects the laser in the opposite direction on a parallel beam path.
In the turning chamber, a chirp-scan setup \cite{lozovoy2008direct, loriot2013self, escoto2018advanced}, including a 5~µm beta-barium borate (BBO) and a spectrometer for the second harmonic, is placed for the temporal characterization of the laser pulses. It measures and compresses the driving laser pulse duration using an acousto-optic programmable dispersive filter (DAZZLER, Fastlite) in the OPS to the Fourier limit just before the HHG source. Fig.~\ref{fig:time} shows a typical temporal structure with a FWHM duration of 4.3~fs, the spectrum, and the spectral phase of the laser pulses.

%%%%%%%%%%%% FIG 2 %%%%%%%%%%%%%%%%%%%%%%%%%
%Fig.~\ref{fig:spectemp}%
%%%%%%%%%%%%%%%%%%%%%%%%%%%%%%%%%%%%%%%%%%%%%%%%%

\begin{figure}
\centering
\includegraphics[width=0.7\linewidth]{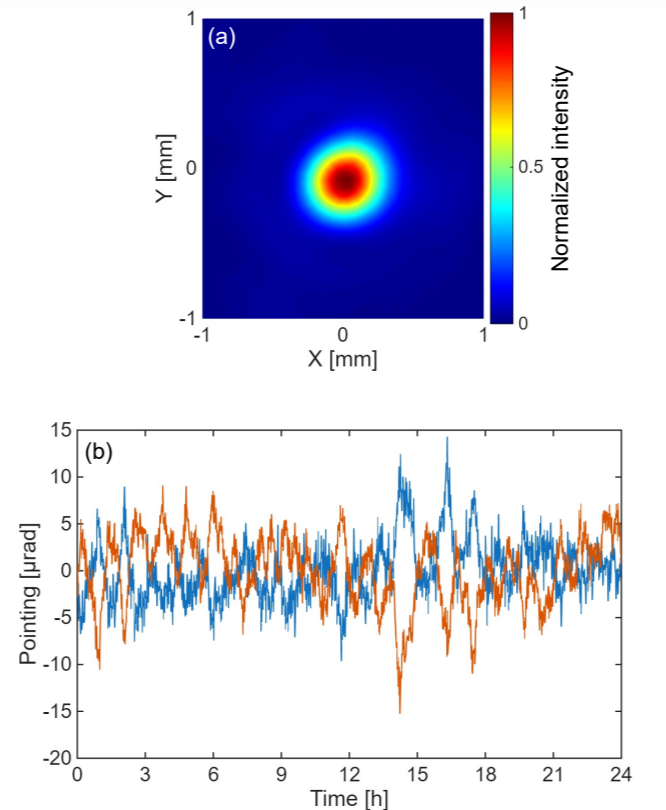}
\caption{\label{fig:lws100focus} (a) Laser focus at the gas target. The FWHM focus size is 425~µm for the applied 22 m focusing. (b) Pointing of the He-Ne laser over 24~h. Red and blue lines represent the horizontal (X) and vertical (Y) pointing. The RMS pointing stabilities in X/Y direction are 3.2/3.7~µrad, and the overall pointing is 4.9~µrad.}
\end{figure}

To make use of the high peak power of the laser, upscaling of HHG in \textcolor{black}{a} gas medium has to be utilized. Maintaining a good phase matching, which is essential to obtain high XUV energies, limits the optimal laser intensity at the gas target. Hence, energy upscaling requires the use of a long focal length to obtain a large focal spot size \cite{heyl2016scale}.
To facilitate a long focal length and adjust the intensity at the gas target, the AS beamline in our laboratory supports a focal length of 22~m and a slightly tunable reduced beam diameter as described before.
The astigmatism introduced by this focusing geometry and the higher-order aberrations are compensated with an adaptive mirror (Imagine Optic) to optimize the energy content in the focal spot. Fig.~\ref{fig:lws100focus}\textcolor{black}{(a)} shows a focal spot measurement of LWS100 after wavefront correction with the adaptive mirror. The FWHM focus size was determined to be 425~µm, and the energy content in the FWHM is 37\%. Furthermore, Fig. ~\ref{fig:lws100focus}\textcolor{black}{(b)} shows the pointing stability of the AS beamline over a time of 24 h, which was measured at the focus with an alignment He-Ne laser that was coupled into the beamline before the compressor of the LWS100 laser system. The pointing over this timescale has a root mean square (RMS) of 4.9~µrad. For a duration of 100~s, the pointing of the He-Ne laser is 1.5~µrad and for LWS100, this increases only to 2.5~µrad, which corresponds to a change in the focus position of only about 10\% of the FWHM spot size.
%%%%%%%%%%%%%%%%%% FIG 3 %%%%%%%%%
%Fig.~\ref{fig:focus}%
%%%%%%%%%%%%%%%%%%%%%%%%%%%%%%%

\begin{figure}
\centering
\includegraphics[width=0.7\linewidth]{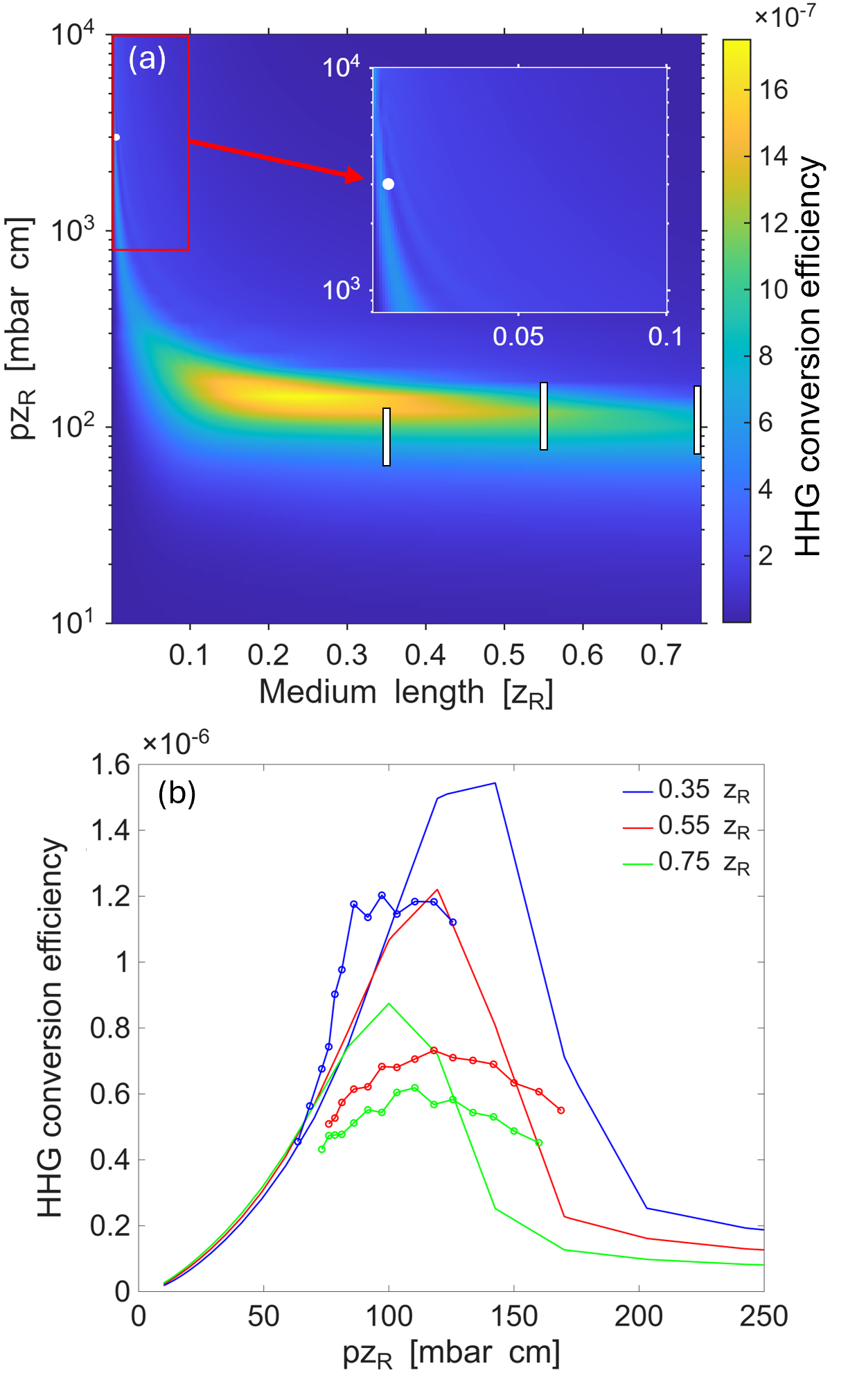}% Here is how to import EPS art
\caption{\label{fig:lund} (a) Simulated (color 2D plot) and measured (3 white rectangles and a dot) HHG conversion efficiency in the Zr spectral window (65-150~eV). p is the neon pressure inside the gas cell and $z_R$  is the Rayleigh length of LWS100 (=~40~cm) with 22~m focusing and (a reduced) 45~mm beam diameter. The simulations are independent of $z_R$, if the efficiency is plotted vs. p$z_R$ and the medium length is shown in units of $z_R$. (b) Simulated (solid lines) and measured (symbols) conversion efficiencies for the 3 rectangular regions in (a). The corresponding maximal HHG conversion efficiencies were $1.2\times10^{-6}$, $7\times10^{-7}$ and $6\times10^{-7}$ for a cell length of 0.35, 0.55 and 0.75$z_R$, correspondingly.
%XUV energy are 42~nJ, 25.6~nJ and 21.7~nJ for a cell length of 0.35, 0.55 and 0.75$z_R$, correspondingly.
}
\end{figure}

Neon is used as the main HHG source with both gas jet and gas cell configurations. To achieve the highest XUV energy, the HHG efficiency for various pressures and gas source sizes were calculated and measured [jet: 2~mm (0.005$z_R$), cell: 14 (0.35$z_R$), 22 (0.55$z_R$) and 30~cm (0.75$z_R$)], as shown in Fig.~\ref{fig:lund}(a) with the measurements in white dot and rectangles.
In Fig.~\ref{fig:lund}(a), the HHG conversion efficiency is plotted (in color) vs. gas pressure and length of the gas cell. $z_R$ is the Rayleigh range of the laser, which was measured to be 40~cm, and p is the gas pressure of the generation medium. 
The pressure of the jet was determined by the applied backing pressure and numerical simulations of the operation of the supersonic nozzle relating this pressure to the pressure at the exit.
For determining p inside the gas cells, a pressure gauge was placed outside the vacuum chamber, which was connected to the cell by a short (8~cm) vacuum pipe. The influence of the connecting pipe length was assessed by extending the pipe length up to 150~cm. The measured pressure exhibited a non-linear relationship with the pipe length, which was addressed through precise evaluations. This measurement was conducted for different gas cell lengths with the same input flow. The pressure in the cell was found to be independent of cell length but strongly dependent on the entrance and exit hole size. 
The length of the gas medium, as well as the target pressure, were optimized to achieve maximal conversion efficiency [Fig.~\ref{fig:lund}(b)], and the experimental optimum was found to be a 14~cm long gas cell with a pressure of approximately 2.6~mbar. During this investigation the experimentally obtained HHG efficiencies in the Zr window were $1.2\times10^{-6}$, $7\times10^{-7}$ and $6\times10^{-7}$ for a cell length of 0.35, 0.55 and 0.75$z_R$, correspondingly. While the jet provided $3\times10^{-7}$ efficiency.

%XUV energies in the Zr window were 42 nJ, 25.6 nJ and 21.7 nJ for a cell length of 0.35, 0.55 and 0.75$z_R$, correspondingly. While the jet provided 10~nJ energy.

The measured conversion efficiency data are compared to simulations with the same input laser parameters using a wide range of pressures and gas source sizes.
The simulated data are obtained by solving the propagation equation for the fundamental and harmonic fields, within the slowly-varying and paraxial approximations. The numerical method presented before\cite{LHuillier1992,Weissenbilder2022} makes use of tabulated single-atom data obtained by solving the time-dependent Schrödinger equation. These simulations include the reshaping of the fundamental field due to partial ionization of the medium. 
A very good agreement was found between the simulations and experiments in terms of absolute conversion efficiency (a deviation of $20\%$ to $40\%$) and pressure dependence. Based on the simulation data in Fig.~\ref{fig:lund}(a), a reduction in the length of the gas cell is likely to enhance the overall efficiency slightly. Experiment with a 7~cm cell did not show, however, higher XUV energy under similar conditions. Therefore, the 14~cm gas cell length was chosen in this  22~m focusing beamline. The optimum was found to be very stable from day to day (see details on energy measurements later).

Argon and helium were tested as HHG generation medium, but neither gas provided optimal conditions for our beamline. In the case of argon, the XUV spectral cutoff was below 70~eV as expected, and achieving optimal HHG necessitated a factor of three further reduction of the laser energy compared to neon, which is an inefficient application of our laser.
Regarding helium, the higher optimal pressure was not reached as it surpassed the vacuum turbo pump’s ability to remove the excess gas. Furthermore, the lower conversion efficiency resulted in significantly lower XUV energies in the Zr window (<1~nJ), which did not support nonlinear studies with the attosecond pulses. 

%%%%%%%%%%%%%%%%%% FIG 4 %%%%%%%%%
%%%%%%%%%%%%%%%%%%%%%%%%%%%%%%%

\section{XUV characterization}

After HHG, the XUV pulses propagate collinearly with the laser 14~m through a vacuum tube towards the XUV characterization and application section of the AS beamline. At the beginning of this section, the XUV beam passes through a holey spherical mirror (mirror with a hole), which reflects a part of the more divergent laser beam onto a CCD camera \textcolor{black}{(not presented in Fig.~\ref{fig:scheme})} near to the generation position. This camera is used for the optimization and observation of the laser alignment at the HHG gas cell, which is carefully imaged onto it. Hence, potential slow drifts of the laser can be easily compensated. Afterwards, the XUV beam is reflected by a Zerodur substrate in a 2.5$^\text{o}$ grazing incidence angle. This substrate reduces the intensity of the laser by about a factor of 2 to avoid damage in the beam path and has >75$\%$ reflectivity in the XUV. Then, the XUV beam is reflected from a gold-coated split mirror in about 2.5$^\text{o}$ grazing incidence angle (a more precise calibration is presented later). Those two gold mirrors act as a beam splitter to enable separation of pump and probe beams. A major advantage of the grazing-incidence geometry, apart from the high reflectivity of the gold mirrors in the XUV ($>\!90\%$ below 150~eV), is the minimal energy loss in the spatial gap between the two mirrors, since the small beam spans several centimeters on their surfaces. For XUV pump-XUV probe experiments, attosecond accuracy is required. To achieve this, one of the two gold mirrors is placed on a piezo actuator (piezosystem jena) with feedback and a range of 75~µm that enables a precise alignment of the time delay between the two XUV pulses. Another translation stage with a larger range is positioned below this piezo actuator to extend the delay range if necessary.
Furthermore, the tilt and tip alignment degrees of this gold mirror are also controlled by two other piezo actuators with feedback to achieve high alignment precision. The stability of this split-and-delay stage is of utmost importance to maintain the high spatial and temporal resolution achieved by the attosecond XUV pulses and requires precise characterization. Over a duration of 1~h, the absolute RMS pointing stability was found to be <3~µrad %OLD:6.5 and 13.6 
for the x direction (horizontal), <6~µrad %OLD:6.4  and 6.9
for the y direction, while the relative pointing difference between pump and probe arms is 2~µrad for the x direction (horizontal), <1.5~µrad for the y direction. The in-loop RMS temporal delay jitter from the position signal of the piezo actuator was measured to be 0.9~as.
%In ellipsoid focus (f=125mm) Pixel resolution (1.4µm) limited, although sub-pix accuracy is reached.
%Evaluation Mohammad:       X       Y           Xpointing       Ypointing
%Stable arm abs RMS (µm)    0.896   0.709       7.2             5.7
%Adjustable arm abs (µm)    0.731   1.014       5.8             8.1
%Average (µrad)             6.5     6.9
%Relative RMS (µm)          0.882   0.418       7.1             3.3

%In focus outside extra focusing mirror (f=750mm) Good resolution. FINAL
%Evaluation Mohammad:       X       Y           Xpointing       Ypointing
%Stable arm abs RMS (µm)    1.664   4.191       2.2             5.59
%Adjustable arm abs (µm)    1.702   3.642       2.27            4.86
%Average (µrad)             2.24     5.23
%Relative RMS (µm)          1.490   1.046       1.99            1.39

\begin{figure}
    \centering
    \includegraphics[width=0.7\linewidth]{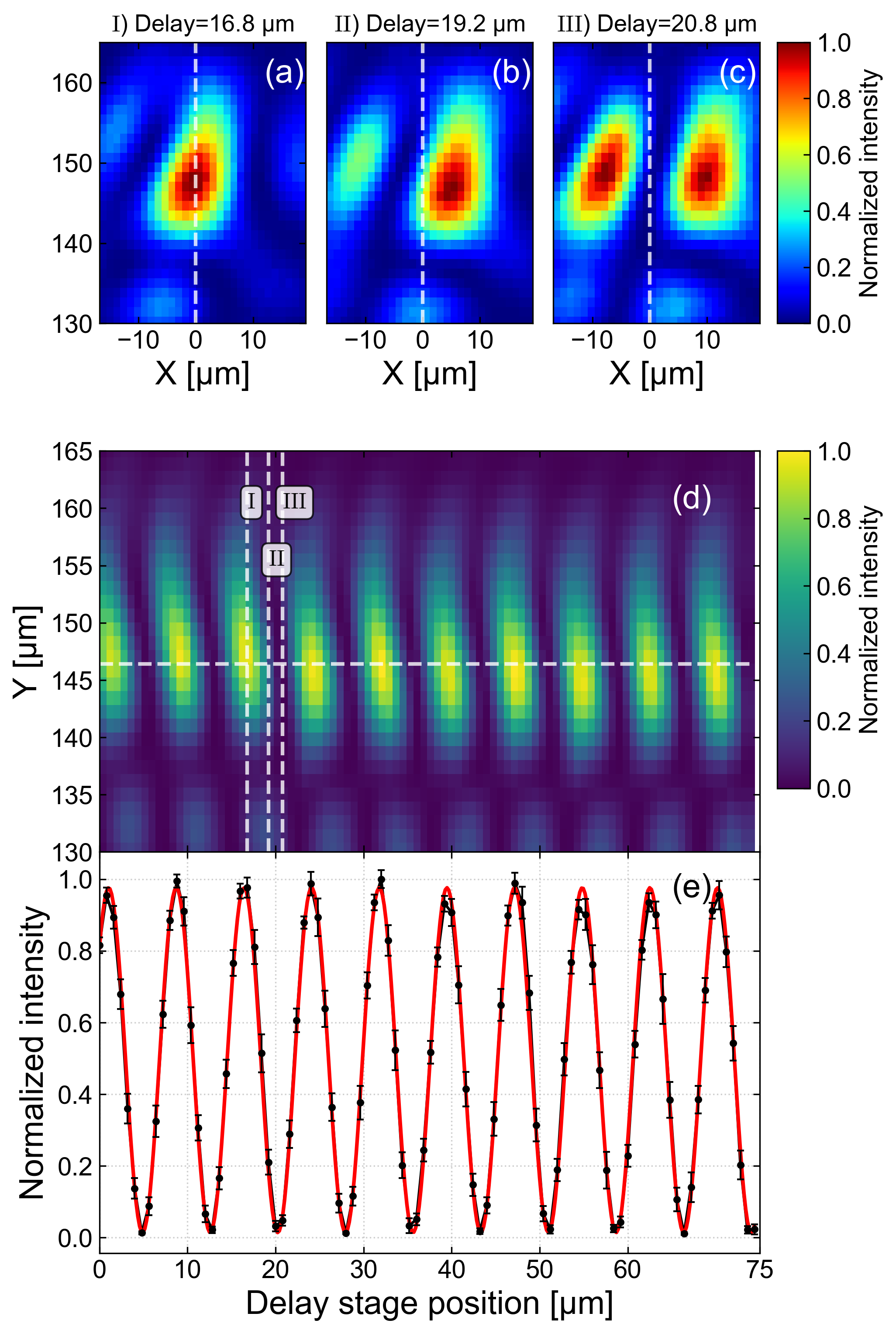}
    \caption{Temporal calibration of the split-and-delay stage and out-of-loop temporal delay jitter measurement. (a-c) Single shot He-Ne focus in the focal plane of the ellipsoidal mirror for three different delays. Dashed lines indicate where a vertical line out is taken. (d) The vertical line out of the He-Ne focus averaged over 50 shots vs. delay stage position. The dashed line indicates where a horizontal line out around the maximum signal is taken. (e) The horizontal line out showing the sinusoidal oscillations. The red line is a sinusoidal fit to the oscillations that provides the temporal calibration factor vs. the stabilized delay stage position.
    This calibration with the stability of the signal around the zero transition of the sinus oscillation (around the value of 0.5) gives an upper limit to the out-of-loop RMS temporal delay jitter.}
    \label{fig:delaycalib}
\end{figure}

The temporal calibration was performed by using a He-Ne laser reflected on both gold mirrors and by measuring its focus in the focal plane of the ellipsoidal mirror (see XUV focusing later). This focus at three different delay between the two halves of 16.8~µm, 19.2~µm, and 20.8~µm, which correspond relative delays of about 0, quarter and half optical period of the He-Ne laser (632.8~nm), are shown in Fig.~\ref{fig:delaycalib}(a-c).
A vertical line out of this focus, indicated as dashed line in Fig.~\ref{fig:delaycalib}(a-c), averaged over 50 \textcolor{black}{measurements}  as a function of the delay stage position of the movable gold mirror is plotted in Fig.~\ref{fig:delaycalib}(d). A horizontal line of this plot is shown in Fig.~\ref{fig:delaycalib}(e), clearly indicating the interference maxima and minima of the He-Ne laser radiation. A sinusoidal fit to these measurement points is shown in red. The obtained period of this fit is $0.8177 \pm 0.0003$~µm$^{-1} = 2\pi\beta/T$, where $T=2109$~as is the optical period and $\beta$ is the temporal calibration factor vs. the stabilized delay stage position in as/µm. The obtained value for this calibration factor is 274.5~$\pm$~0.1~as/µm, which gives a full delay range of 20.6~fs using the 75~µm range of the delay stage.
This provides, beyond the calibration, the grazing incidence angle on the gold mirrors ($\alpha$) utilizing the $\tau = 2\sin(\alpha)x/c$ relationship between the delay ($\tau$) and the delay stage position, i.e., shift of the movable gold mirror ($x$). It delivers $\alpha = 2.36^{\circ}$. Furthermore, the stability of the signal at the zero transition of the sinus oscillation, i.e., around 0.5 in Fig.~\ref{fig:delaycalib}(e), provides an upper limit to the out-of-loop RMS temporal delay jitter of 20-25 as.

A potential drawback of this grazing incidence geometry of the split-and-delay stage is the transverse position change of the XUV focus with the delay, i.e., the delay stage position. This was measured with the He-Ne laser for the full delay range of >20~fs, and a transverse shift of less than 1~µm has been obtained, which was limited by the pointing stability. This shift corresponds only to 10-15\% of the XUV focus size. This low value of the shift is originating from the \textcolor{black}{large demagnification of the ellipsoidal XUV mirror, which is much higher than other designs\cite{Campi2016}.}
%the optimal geometry with perpendicular planes for the split-and-delay unit and the focusing, which is different from other designs\cite{Campi2016}.

%%%%%%%%%%%%%%%%%%%%%%%%%%%%%%%%%%%%%%%%%%%%Fig%%%%%%%%%%%%%%%%%%%%%%%%%%%%%%%
%fit function A * np.sin(B * x + C) + D
%Amplitude (A): 0.4810 ± 0.0035
%Offset (D): 0.495 ± 0.002
%Offset (B): 0.8177 ± 0.0003
%R-squared: 0.9952
%
% Period of sinus: 0.8177 ± 0.0003 µm^-1 = 2*pi*beta/T, where T = 2.109 fs period of He-Ne laser, beta[fs/µm] calibrtation factor that we want to determine.
%%%%%%%%%%%%%%%%%%%%%%%%%%%%%%%%%%%%%%%%%%%%%%%%%%%%%%%%%%%%%%%%%%%%%%%%%%%%%%

To remove the remainder of the fundamental VIS-NIR laser light, the XUV beam passes through several thin-film metallic filters after the split-and-delay unit. These filters are mounted on motorized filter wheels, which allow the user to choose the desired spectral region \cite{HENKE1993181}.
Behind these spectral filters, the XUV beam can be sent to a calibrated XUV photodiode for measuring the pulse energy, an XUV CCD camera for measuring the beam profile, and an XUV flat-field spectrometer for the characterization of the spectrum. The photodiode can measure single-shot pulse energy, which is crucial for the rapid optimization of the HHG process. Furthermore, it allows the measurement of the shot-to-shot energy fluctuations of the XUV pulses. In addition, the XUV photodiode enables the characterization of the transmission of the metallic filters, which degrades continuously over the course of every few months due to oxidation. The experimentally verified transmission values during the presented energy measurements are shown in Table 1. In this experiment, different combinations of filters from the same material were used, and the transmission of each was evaluated, assuming that XUV energy and laser energy remained constant during the measurement. It has to be mentioned that the measured values are lower than the literature values for comparable filters, which is commonly caused by oxidation of the filter surface \cite{HENKE1993181}. To maximize the attosecond pulse energy, the position of the entrance and exit of the gas cell, as well as the wavefront and compression of the driving laser at the gas target, are fine-tuned with the XUV signal on a daily basis. Typically the photodiode provides the feedback signal for this procedure.

%%%%%%%%%%%%%% TABLE 1 %%%%%%%%%%%%%%%%%%%%%%%%%%%%%%%%%%%%%%%%%%%%%%%%%%%%

\begin{table}[ht]
\caption{\label{tab:table4}The different metal filters of the AS beamline and their spectral ranges and transmissions.}
\centering
\begin{threeparttable}
\begin{tabular}{|l|l|l|l|}
\hline
Filter & Thickness (nm) & Spectral range & Transmission \\
\hline
Zr 1 & 150 & $\sim$ 65 eV-C\tnote{a} & $38 \pm 2\%$ \\
Zr 2 & 150 & $\sim$ 65 eV-C & $45 \pm 2\%$ \\
Zr 3 & 150 & $\sim$ 65 eV-C & $48 \pm 2\%$ \\
\hline
Pd & 100 & $\sim$ 95 eV-C & $14 \pm 1\%$\tnote{b} \\
\hline
Al 1 & 500 & $\sim$ 18 eV–73 eV & $28\%$ \\
Al 2 & 500 & $\sim$ 18 eV–73 eV & $16\%$ \\
\hline
Si & 500 & $\sim$ 20 eV–100 eV & Not measured \\
\hline
\end{tabular}

\begin{tablenotes}
\item[a] Cutoff.
\item[b] Measured average value for the spectral region of the Zr window, including the range where Pd does not transmit.
\end{tablenotes}
\end{threeparttable}

\end{table}
%%%%%%%%%%%%%%%%%%%%%%%%%%%%%%%%%%%%%%%%%%%%%%%%%%%%%%%%%%%%

\begin{figure}
\centering
\includegraphics[width=0.7\linewidth]{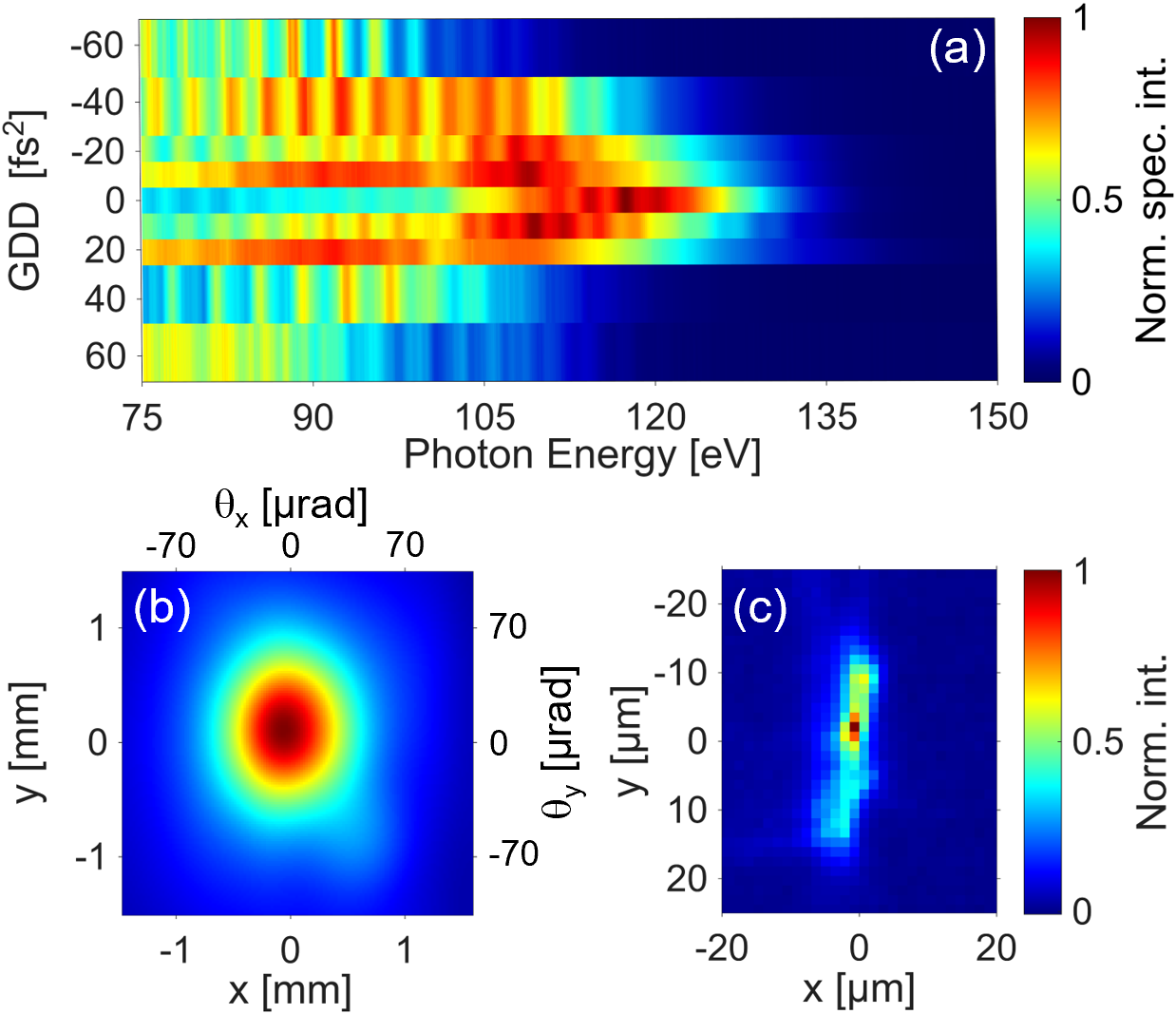}% Here is how to import EPS art
\caption{\label{fig:gdd} (a) HHG spectrum vs. GDD values of LWS100, where 0~fs\textsuperscript{2} corresponds to the optimal compression. (b) Single-shot XUV beam profile 14~m after generation. The average FWHM beam diameter is 1.03~mm, which corresponds to a beam divergence of 73~µrad. (c) 50 single-shot XUV foci are averaged, leading to an average FWHM of 5.7~µm.}
\end{figure}

If the isolation of the attosecond pulses is important for a specific experiment, the compression of the laser is fine-tuned by the use of the Dazzler as a control unit and the XUV spectrometer as a feedback signal to increase the continuum part of the spectrum. The third and fourth order spectral phase are set to the optimal value for compression, and the group-delay dispersion (GDD) is changed using the Dazzler, which ensures very good temporal compression at the gas target. Measured spectra vs. GDD is shown in Fig.~\ref{fig:gdd}\textcolor{black}{(a)}. In this figure, each spectrum is measured with a Zr filter and is an average of 50 shots. The spectra were measured with 10~fs\textsuperscript{2} GDD steps around the optimal compression (highest cutoff), and the step size was increased to 20~fs\textsuperscript{2} further away. Typically, no or no significant GDD change was needed to reach the best continuum.

%%%%%%%%%%%%%%%%%% FIG 5 %%%%%%%%%
%Fig.~\ref{fig:GDD}%
%%%%%%%%%%%%%%%%%%%%%%%%%%%%%%%%%%%%%%%%%%%%%%%%%%%%%%%%%%%%%%%%%%%%%%%%%%%%

With accurate alignment of HHG and laser parameters, XUV pulse energies up to 55.0~±~5.2~nJ in the Zr spectral window could be achieved, taking into account the transmission of the filters. Furthermore, pulse energies above 40~nJ, corresponding to an on-target energy of above 10~nJ, are commonly reached with a typical RMS energy stability of 5-10\%. \textcolor{black}{With a measured focal spot size of approximately 6~\textmu m (FWHM) and a pulse duration estimated from the spectrum and realistic chirp of about 200~as, the resulting peak XUV intensity reaches approximately $10^{14}$~W/cm$^2$.} Those energies are twice as high as previously published\cite{bergues2018tabletop} and >100 times higher than the energy of common kHz HHG beamlines\cite{Osolodkov2020,Lorek20214,Rothhardt2016}. The relatively low RMS error further minimizes the measurement uncertainties for investigations with this XUV source.

Investigations of the beam profile with the XUV CCD camera reveal that the XUV beam is very well collimated. 14~m behind the gas source the measured beam size varies between 0.7 to 1.5~mm, depending on the focal spot size of the driving laser. A typical beam profile on both spatial and angular x- and y-scales is shown in Fig.~\ref{fig:gdd}, applying the optimal laser spot size of 425~µm for the highest XUV energy. The average FWHM beam size is about 1~mm, which corresponds to a beam divergence of 70~µrad. Furthermore, non-circular XUV beams can be produced and characterized when the driving laser is slightly astigmatic (not visible on the laser focus) due to misalignment of the deformable mirror. However, this leads to a significant reduction in conversion efficiency.

\begin{figure}
\centering
\includegraphics[width=0.7\linewidth]{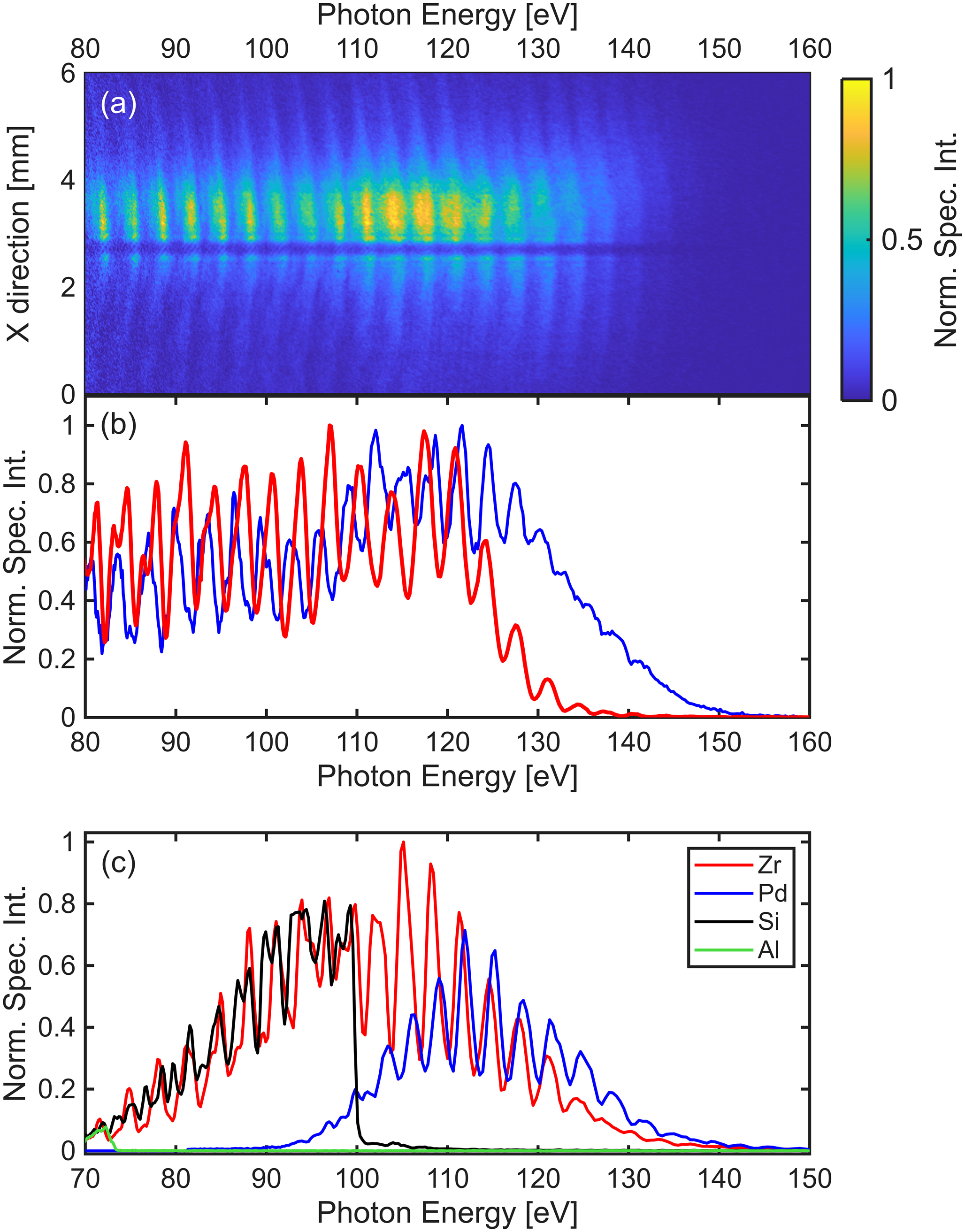}% Here is how to import EPS art
\caption{\label{fig:xuvspec} (a) Single-shot 2D XUV spectrum. (b) Single-shot XUV spectrum for different CEP values. (c) XUV spectrum behind Zr, Pd, Si, and Al thin film metallic filters \textcolor{black}{without CEP stabilization}.}
\end{figure}

The XUV flat-field spectrometer of the AS beamline measures the spectrum as a function of one transverse coordinate (x). One such measurement is shown in Fig.~\ref{fig:xuvspec}\textcolor{black}{(a), indicating a homogeneous spectral structure with the transverse (X) position.} \textcolor{black}{The dark line in this figure at around X=2.7~mm originates from the gold split mirror used to separate the XUV pump and XUV probe beams. With a motorized mirror located in the vacuum chamber that turns the laser backwards, the position of the XUV beam on these two mirrors can be adjusted. Near this dark area at the top and bottom, an interference pattern is noticeable due to the scattering caused by the edges of the split mirror}. Spectral measurements in Fig.~\ref{fig:xuvspec}\textcolor{black}{(b)} show a cutoff photon energy of approximately 150 eV. Furthermore, for specific carrier-envelope phase (CEP) values, the harmonic lines in the cutoff region vanish and form a continuum, indicating that the XUV pulses generated in this spectral region are isolated. One XUV spectrum with (blue) and one without (red) continuum is shown in Fig. \ref{fig:xuvspec}\textcolor{black}{(b)}. Fig. \ref{fig:xuvspec}\textcolor{black}{(c)} shows the spectral tunability of the source with different filters that are typically used in the filter wheels \textcolor{black}{(CEP was not stabilized during the measurement with filters)}. The Zr spectrum (blue) is the standard because it provides the highest photon energy and the highest intensity on target, which is crucial for nonlinear experiments. The Pd filter (red) is used when high photon energies or good temporal isolation of the attosecond pulses are important for an experiment. The Al and Si filters are used for the calibration of the XUV spectrometer. Moreover, the Si filter is also used to suppress higher photon energies >100 eV. This is necessary for certain experiments, e.g., the nonlinear 4x ionization of Xenon (Xe → Xe$^{4+}$), which would be masked by linear ionization triggered by photons originating from the cutoff spectral region\cite{bergues2018tabletop}.

%%%%%%%%%%%%%%%%%%% FIG 6 %%%%%%%%%%%%%%%%%%%%%%%%%%%%%%%%%%%%%%%%%%%%%%%%
%Fig.~\ref{fig:XUVspectra}%
%%%%%%%%%%%%%%%%%%%%%%%%%%%%%%%%%%%%%%%%%%%%%%%%%%%%%%%%%%%%%%%%%%%%%%%%%%%%
%%%%%%%%%%%% FIG 7 %%%%%%%%%%%%%%%%%%%%%%%%%
%Fig.~\ref{fig:spectemp}%
\begin{figure*}[!t]
\includegraphics[width=\linewidth]{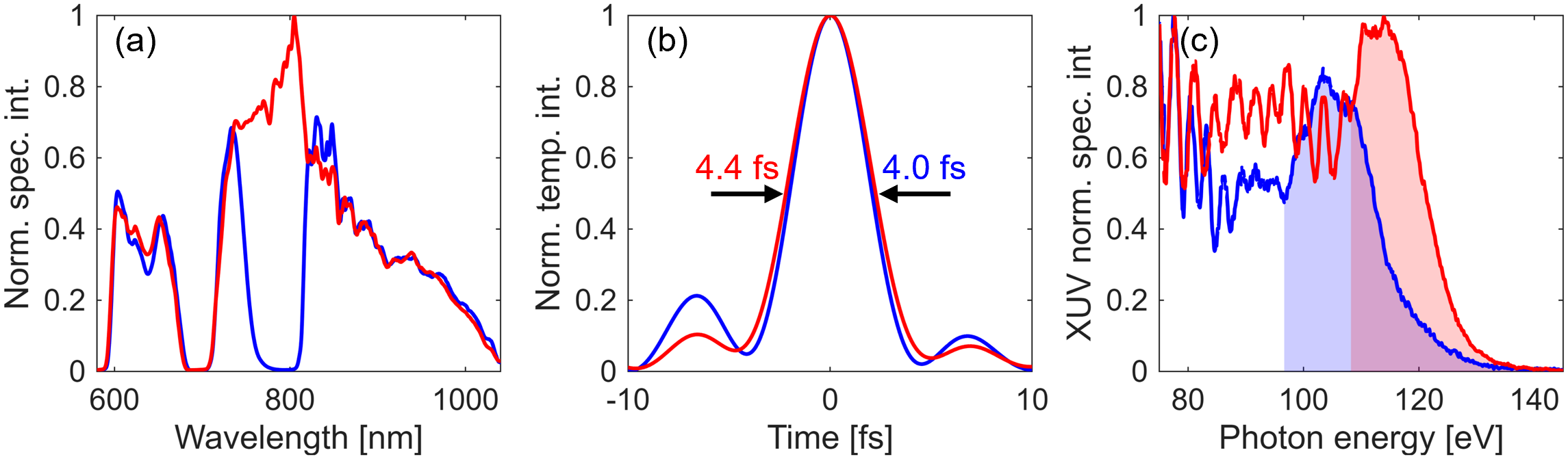}% Here is how to import EPS art
\caption{\label{fig:spectemp} Spectral amplitude modulation for temporal super-resolution. (a) Spectrum of the laser before (red) and after (blue) amplitude modulation. (b) Temporal intensity of the laser before (red) and after (blue) amplitude modulation. (c) HHG XUV spectrum without (red) and with (blue) amplitude modulation.}
\end{figure*}
%%%%%%%%%%%%%%%%%%%%%%%%%%%%%%%%%%%%%%%%%%%%%%%%%
%%%%%%%%%%%%%%%%%%% FIG 8 %%%%%%%%%%%%%%%%%%%%%%%%%%%%%%%%%%%%%%%%%%%%%%%%
%Fig.~\ref{fig:ion}%
\begin{figure}[!t]
\centering
\includegraphics[width=0.7\linewidth]{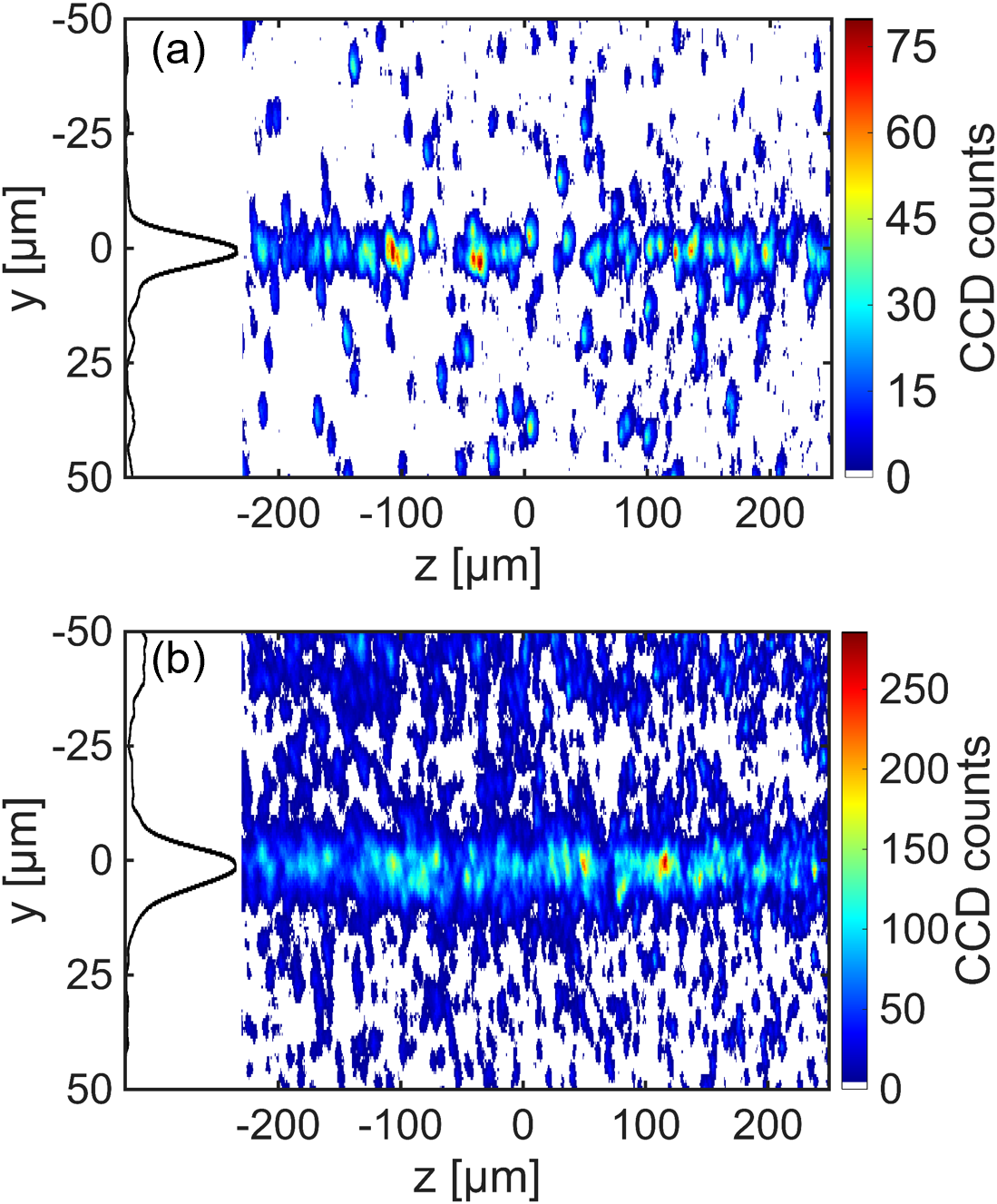}% Here is how to import EPS art
\caption{\label{fig:ion} Ion distribution of linearly ionized Xe species measured with the ion microscope. (a) Focusing with the spherical back-reflecting multilayer mirror. (b) Focusing with the ellipsoidal grazing incidence mirror. For both measurements, 10 images with 2 s exposure time (200 XUV pulses) were averaged.}
\end{figure} 
%%%%%%%%%%%%%%%%%%%%%%%%%%%%%%%%%%%%%%%%%%%%%%%%%%%%%%%%%%%%%%%%%%%%%%%%%%%%

\section{Temporal super-resolution}
Extending the spectral continuum is highly beneficial, as it indicates the generation of increasingly well-isolated and more energetic attosecond pulses\cite{rivas2016generation}. To this end, temporal super-resolution via spectral amplitude modulation serves as a key technique that narrows the temporal FWHM of the driver laser pulses\cite{temporalsupperresolution}. Figures~\ref{fig:spectemp}(a) and (b) illustrate this technique. By cutting the laser spectrum around its central wavelength, a reduction in FWHM pulse duration by 400~as, down to 4.0~fs, is achievable, while maintaining the same spectral range and central wavelength. This reduction in FWHM subsequently leads to a measured 35\% expansion of the continuum part of the XUV spectrum, as depicted in Fig.~\ref{fig:spectemp}\textcolor{black}{(c)}. Although this approach leads to a 28\% reduction in driving laser intensity within the gas cell, it is easily compensated for by increasing the beam diameter using an iris before the AS beamline. With a larger beam diameter, the focal spot size decreases, reducing the effective area for HHG and ultimately leading to an XUV energy reduction of approximately 10\%. Both of the presented XUV spectra were measured in the saturated HHG regime \cite{rivas2016generation} to produce a broader continuum region. Temporal super-resolution can produce even shorter pulses, down to 3.7~fs \cite{lws100}, and thus better isolation, but at the cost of a further reduction in XUV energy.

%%%%%%%%%%%%%% FIG 9 %%%%%%%%%%%%%%%%%%%%%%%%%%%%%%%%%%%%%%%%%%%%
%Fig.~\ref{fig:VMI}%
%\begin{figure*}
%\includegraphics[width=\linewidth]{Fig 10_V4.png}% Here is how to import EPS art
%\caption{\label{fig:VMI}VMI design and simulations for electron energy resolution.
%(a) Design and simulated electron trajectories for kinetic energies from 10~eV to 240~eV. (b) Energy resolution (<3~eV for the entire energy range), $\Delta$E/E <0.01 around 200~eV electron energy. (c) Simulation of detected signal with electron energies from 100~eV to 240~eV with 20~eV steps (left) and its Abel inversion (right) for 3D momentum reconstruction.}
%\end{figure*}
%%%%%%%%%%%%%%%%%%%%%%%%%%%%%%%%%%%%%%%%%%%%%%%%%%%%%%%%%%%%%%%%%
\section{XUV focusing and applications}
As shown in Fig. \ref{fig:scheme} if the XUV beam is not sent to one of the three detectors, it impinges on a gold-coated ellipsoidal mirror, which focuses the XUV beam with a focal length of 125~mm onto a CMOS camera or into the field of view of an ion microscope. The ellipsoidal mirror has a grazing incidence angle of 8$^\text{o}$, it supports the entire spectral range of the Zr window and has a peak reflectivity of 75\%. The CMOS camera is used for the characterization and optimization of the XUV focus\cite{Muschet2022}. Its ability to acquire 2D images of the focus with only a few and even a single shot is crucial because it enables rapid alignment of the grazing incidence focusing mirror, that has six degrees of freedom and is therefore difficult to align\cite{bourassin2013focus}. The feedback from the CMOS camera is used to maximize the peak intensity of the XUV beam during the alignment procedure of the ellipsoidal mirror. Several axes of the mirror are coupled to each other, which makes 2D scans during the alignment necessary. Fig.~\ref{fig:gdd}\textcolor{black}{(c)} shows an average of 50 single-shots XUV foci with an average FWHM of 5.7~µm. 

Smearing effects due to the non-collinear geometry in the focus reduce the temporal resolution. This turned out to be the dominant limitation to the measurement accuracy, estimated to 
%The temporal resolution due to this geometric smearing at different positions in the foci of the two XUV beams is 
about 50~as (with the above focus size and an angle of 4~mrad between the XUV beams). It can be improved by reducing the beam diameter in front of the ellipsoidal focusing mirror. 
%Originally 1 mm XUV beam diameter that is split into two --> 0.5 mm (more correctly with half circe beams 0.42 mm) distance between them and 125 mm focal length --> 4 mrad (more correctly 3.4 mrad), smearing (FWHM 6µm focus) is 2*4mrad*6µm/2/0.3µm/fs/sqrt(2) = 57 as (more correctly 48 as). The 2* is due to the + and - delay in the AC, the 4mrad angle is full angle and 6µm/2 is the HWHM radius, the +- smearing (x2 should be larger) is compensated by the fullangle, which is (x2 larger than needed). The sqrt(2) corresponds to the FWHM in the 2nd order nonlinear signal for a Gaussian beam as at the HM I=I0/2 and signal is ~I^2. 0.8 mm beam diamter (Alex XUV focus) gives 40 as smearing. Correctly the signal should be integrated in space with different delays, but we do not do it.
% Resolution in Tzallas paper (Journal of modern optics 52, 321–340 (2005)): 4 mm beam diameter, 150 mm focal length, 2w = 10 µm (FWHM 6 µm) --> 2/150 = 13.3 mrad --> 190 as resolution (more correctly 160 as)
%
In addition to grazing incidence focusing, this beamline also offers the possibility to focus the beam with spherical back-reflecting multilayer mirrors having also 125~mm focal length \textcolor{black}{(not presented in Fig.~\ref{fig:scheme})}. These allow investigations with a very small focus size down to 1~µm \cite{bergues2018tabletop}. Furthermore, the spectral region of the focused beam is tunable over a wide range\cite{haelbich1976multilayer, hofstetter2011lanthanum}. However, the spectral bandwidth of back-reflecting mirrors is limited, and the energy throughput is drastically reduced, e.g., 15\% peak reflectivity with 10~eV bandwidth at 93~eV photon energy.

The ion microscope acquires a magnified image of the (x) projected ion distribution that is generated by the XUV pulse in a low density gas\cite{tsatrafyllis2016ion}. It consists of a repeller and an extractor, between which the low density gas sample is placed and where the XUV beam is focused. It also includes two electrostatic lenses, which project a magnified image of the ion distribution in the interaction plane onto the detector. Furthermore, the ion microscope has two cylindrical electrostatic lenses, which compensate astigmatism in the image. The detector of the ion microscope is a double microchannel plate (MCP) with a chevron configuration and a phosphor screen. The image of the phosphor is then recorded with a low-noise camera\cite{schultze2011spatially}. Depending on the used geometry (repeller and extractor distance) of the ion microscope, this detector achieves a spatial resolution of 1 or 5~µm corresponding to 2 or 11~mm distance. Moreover, it is possible to gate the MCP with a temporal window $\geq8$~ns. This makes it possible to record the Time Of Flight (TOF) spectrum, which in turn permits quantitative analysis of individual ionization states\cite{tzallas2018time} and their spatial distribution.
Fig. \ref{fig:ion} show the projection of linearly ionized xenon ion distribution (sum of Xe$^{+}$, Xe$^{2+}$, and Xe$^{3+}$) along the x-axis for the spherical back-reflecting mirror (reflectivity: R=15\%, 88-98~eV)  and the ellipsoidal grazing incidence focusing mirror (R=75\%, 65-150~eV) respectively. The obtained focus size with the spherical mirror, measured by the ion microscope of 5.2~µm. It is limited by the resolution of the microscope used with the larger repeller-extractor distance (11~mm), to allow the incoming beam to pass unclipped through the device. The real focus size with the spherical mirror is estimated to be 1 to 2~µm, using the combined observation of linearly and nonlinearly generated ions and careful evaluation\cite{bergues2018tabletop}. The ions in the background are produced by the unfocused incoming XUV beam, which passes through the field of view of the ion microscope in back-reflection focusing geometry.

For direct comparison, the focus of the grazing incidence focusing geometry (Fig.~\ref{fig:ion}\textcolor{black}{(b)}) was measured with the same settings. The size of the projected focus increases to 8.7~µm, which after deconvolution gives 6-7~µm FWHM spot size. This is caused by aberrations from the ellipsoidal mirror, which are especially high along the y-axis. This is clearly shown by the 2D images of the XUV focus, which are acquired with the CMOS camera (Fig.~\ref{fig:gdd}\textcolor{black}{(c)}). The major advantage of the grazing incidence focusing is the broad bandwidth and high reflectivity, which together provide a much higher energy throughput, estimated to be about 27 times that achieved with spherical focusing, for the Zr spectral window. Integrating the plots in Fig.~\ref{fig:ion} indicates about 13 times more measured signal, and thus XUV photons, with the ellipsoidal mirror. This combination of a high-energy XUV and soft X-ray source with high throughput focusing and the ion microscope as a detector, allows for the observation of a manifold of ionization phenomena, like, e.g. single-photon multiple ionization and multi-photon multiple ionization of atomic as well as molecular gases \cite{bergues2018tabletop, kolliopoulos2014single, tsatrafyllis2016ion,Manschwetus2016}.

%\textcolor{black}{
%To complement the spatial diagnostics provided by the ion microscope, the AS beamline is equipped with a dedicated Velocity Map Imaging (VMI) spectrometer\cite{Eppink1997}. This detector is a permanent endstation designed to enable 3D momentum reconstruction of photoelectrons emitted during high-intensity XUV-matter interactions\cite{Rading2018}. The spectrometer is based on a Wiley-McLaren TOF geometry with a 223~mm flight tube, specifically optimized for the high photon energies (up to 150 eV) provided by this source.
%}

In addition to the ion microscope, the AS beamline also includes a Velocity Map Imaging (VMI) detector\cite{Eppink1997}, enabling the measurement of the transverse momentum of electrons \textcolor{black}{ up to 240~eV energy emitted after ionization. Comprehensive design specifications and SIMION simulations demonstrating the electron energy resolution of this spectrometer are provided in the Supplementary Material.}

\section{Summary}
In this paper, we described a high-energy attosecond XUV and soft X-ray source and the associated beamline. The source delivers an energy of 40-50~nJ in the spectral range of 65-150~eV with a spectrum supporting isolated attosecond pulses. The HHG is optimized using both numerical analysis and experimental techniques. The influence of important parameters is studied, such as gas type and pressure, driving laser intensity, focus, and pulse duration, and the best parameters are used to produce stable, isolated, high-energy XUV pulses. Various relevant characteristics of the XUV source are measured, including energy, spectrum, beam profile, and divergence. Different XUV focusing optics are tested, and the XUV focus in the interaction area is carefully measured. To characterize the interaction of matter with the intense attosecond pulses, two detectors are utilized, an ion microscope and an electron VMI spectrometer.
Because of the damage threshold limitation on commercial flat mirrors by the laser, we are constrained to using less than half of the LWS100 energy for HHG. Therefore, temporal super-resolution was also tested, which increases the continuum spectral range and thus the isolation and energy content of the attosecond pulses.
Our experimental findings indicate that HHG can produce intense ultrashort pulses with the necessary intensity levels for conducting nonlinear experiments in the XUV and soft X-ray spectral range.

\section*{Supplementary Material}

See the supplementary material for a detailed technical description of the Velocity Map Imaging (VMI) spectrometer, including SIMION simulations of the electron trajectories, energy resolution benchmarks ($\Delta E/E$), and the Abel inversion procedure used for 3D momentum reconstruction.
\section*{Credit lines}
This article has been submitted to APL Photonics. After it is published, it will be found at https://pubs.aip.org/aip/app.
Copyright © 2026 S. Vardast et al. This article is distributed under a Creative Commons Attribution-NonCommercial-NoDerivs 4.0 International (CC BY-NC-ND) License.

\section*{Acknowledgements}

We acknowledge the support from Vetenskapsrådet (grant nos. \textcolor{black}{2017-04106}, 2019-02376, 2020-05111, \textcolor{black}{2021-05992}, \textcolor{black}{2023-04684} and 2023-04603), Knut och Alice Wallenbergs Stiftelse (grant nos. 2019.0140, \textcolor{black}{2020.0111}, 2024.0120 and 2023.0207). L.V. expresses gratitude for the grants from Kempestiftelserna (grant no. SMK21-0017, JCSMK24-539). A.L. acknowledges the support from Knut och Alice Wallenbergs Stiftelse through the Wallenberg Center for Quantum Technology.

\section*{Data Availability}

The data that support the findings of this study are available from the corresponding author upon reasonable request.

\section*{Author contributions statement}
The research was initiated and led by L.V. The experiment was planned, designed, and performed by S.V., A.M., N.S., and L.V., with contributions from M.R.-P. and F.S. HHG simulations were conducted by R.W. and E.A. under the supervision of A.L. VMI spectrometer simulations were performed by N.S. with contributions from J.L., S.M., P.E.-J., and L.V. All co-authors discussed the results and contributed to the writing of the paper. All authors reviewed the manuscript. 

\section*{Author Declarations}
The authors have no conflicts to disclose.
\newpage
\clearpage
\section*{Supplementary Material: Velocity Map Imaging}

\begin{figure}[!ht]
\includegraphics[width=1\linewidth]{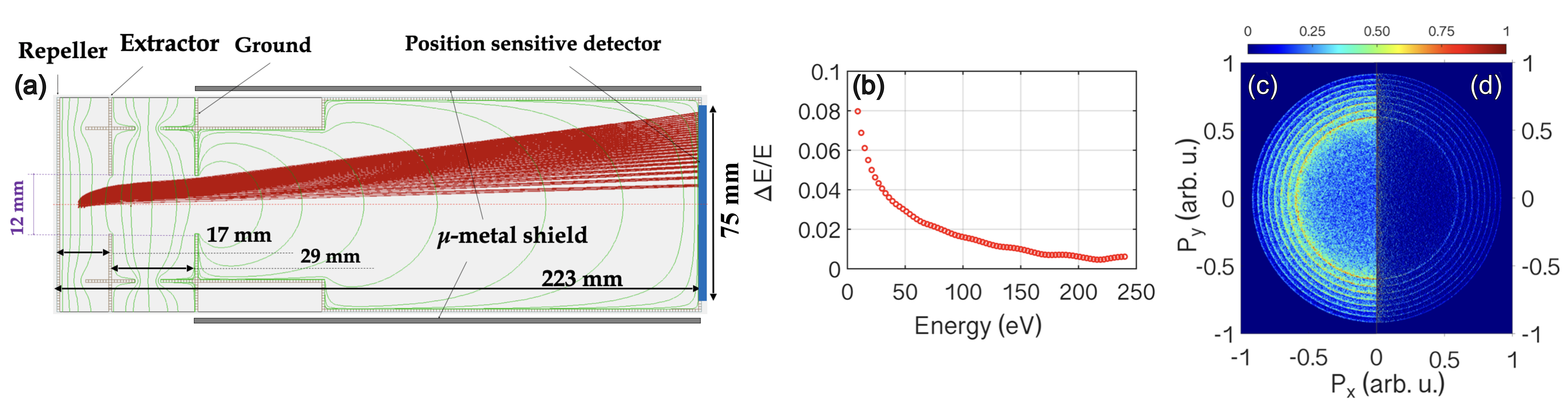}% Here is how to import EPS art
\caption{\label{fig:VMI}VMI design and simulations for electron energy resolution.
(a) Design and simulated electron trajectories for kinetic energies from 10~eV to 240~eV. (b) Energy resolution (<3~eV for the entire energy range), $\Delta$E/E <0.01 around 200~eV electron energy. Simulation of detected signal with electron energies from 100~eV to 240~eV with 20~eV steps (c) and its Abel inversion (d) for 3D momentum reconstruction.}
\end{figure}

In addition to the ion microscope, the AS beamline also includes a Velocity Map Imaging (VMI) detector\cite{Eppink1997}, enabling the measurement of the transverse momentum of electrons emitted after ionization. A dedicated VMI spectrometer is constructed, inspired by the geometry of a former device\cite{Rading2018}, based on the Wiley-McLaren TOF geometry with open electrodes, with a length of 223~mm. The schematic is given in Fig. \ref{fig:VMI}(a). %The separation between the repeller and the extractor is 17~mm and the distance between the extractor and the ground is 29~mm with a 12~mm hole in the center. 
The 177~mm length of the fly tube leads to a resolution better than 3 eV at an electron energy of around 200~eV, so that, consecutive odd harmonics are resolvable.
%Fig.~\ref{fig:VMI}%

A SIMION simulation\cite{dahl2000simion} confirms the required energy resolution, Using 27 particles in a 3$\times$3$\times$3 (of size 1~mm$\times$0.05~mm$\times$0.05~mm) cell 
indicating a $\Delta$E/E better than 0.01 around 200 eV electron energy, as shown in Fig.~\ref{fig:VMI}(b).
In order to evaluate the effect of the detector and hence the 2D projections of the transverse momentum distributions,further particle tracking simulations were performed with $10^6$ photo-electrons for each energies with a momentum vector along the direction of polarization of the laser. A voltage ratio of about 0.72 between extractor and repeller ($V_E/V_R$) is used in the simulation. Energies ranging from 100~eV to 240~eV with a 20~eV step are used to simulate the detected 2D signal, as shown in Fig.~\ref{fig:VMI}(c). The kinetic energy of the electrons is reconstructed from the 2D image by employing Abel inversion, as seen in Fig. \ref{fig:VMI}(d). These simulations unambiguously confirm that measurements with a resolution better than 3~eV in the full electron energy range are possible with such a detector upon implementation in the AS beamline. The VMI is expected to shed light on non-sequential nonlinear ionization processes at electron energies around 200~eV, thereby enabling complete characterization of the attosecond pulse and enhancing the beamline’s capabilities.

%\bibliography{unsrt}
\bibliographystyle{ieeetr}

\bibliography{sample}

@PREAMBLE{
 "\providecommand{\noopsort}[1]{}" 
 # "\providecommand{\singleletter}[1]{#1}%" 
}

@article{krausz2009attosecond,
  title={Attosecond physics},
  author={Krausz, Ferenc and Ivanov, Misha},
  journal={Reviews of Modern Physics},
  volume={81},
  number={1},
  pages={163},
  year={2009},
  publisher={APS}
}

@article{ferray1988multiple,
  title={Multiple-harmonic conversion of 1064 nm radiation in rare gases},
  author={Ferray, M and L'Huillier, Anne and Li, XF and Lompre, LA and Mainfray, G and Manus, C},
  journal={Journal of Physics B: Atomic, Molecular and Optical Physics},
  volume={21},
  number={3},
  pages={L31},
  year={1988},
  publisher={IOP Publishing}
}

@article{mcpherson1987studies,
  title={Studies of multiphoton production of vacuum-ultraviolet radiation in the rare gases},
  author={McPherson, A and Gibson, G and Jara, H and Johann, U and Luk, Ting S and McIntyre, IA and Boyer, Keith and Rhodes, Charles K},
  journal={JOSA B},
  volume={4},
  number={4},
  pages={595--601},
  year={1987},
  publisher={Optica Publishing Group}
}

@article{gaumnitz2017streaking,
  title={Streaking of 43-attosecond soft-X-ray pulses generated by a passively CEP-stable mid-infrared driver},
  author={Gaumnitz, Thomas and Jain, Arohi and Pertot, Yoann and Huppert, Martin and Jordan, Inga and Ardana-Lamas, Fernando and W{\"o}rner, Hans Jakob},
  journal={Optics Express},
  volume={25},
  number={22},
  pages={27506--27518},
  year={2017},
  publisher={Optica Publishing Group}
}

@article{li2020attosecond,
  title={Attosecond science based on high harmonic generation from gases and solids},
  author={Li, Jie and Lu, Jian and Chew, Andrew and Han, Seunghwoi and Li, Jialin and Wu, Yi and Wang, He and Ghimire, Shambhu and Chang, Zenghu},
  journal={Nature Communications},
  volume={11},
  number={1},
  pages={2748},
  year={2020},
  publisher={Nature Publishing Group UK London}
}

@article{sansone2011high,
  title={High-energy attosecond light sources},
  author={Sansone, Giuseppe and Poletto, Luca and Nisoli, Mauro},
  journal={Nature Photonics},
  volume={5},
  number={11},
  pages={655--663},
  year={2011},
  publisher={Nature Publishing Group UK London}
}

@article{duris2020tunable,
  title={Tunable isolated attosecond X-ray pulses with gigawatt peak power from a free-electron laser},
  author={Duris, Joseph and Li, Siqi and Driver, Taran and Champenois, Elio G and MacArthur, James P and Lutman, Alberto A and Zhang, Zhen and Rosenberger, Philipp and Aldrich, Jeff W and Coffee, Ryan and others},
  journal={Nature Photonics},
  volume={14},
  number={1},
  pages={30--36},
  year={2020},
  publisher={Nature Publishing Group UK London}
}

@article{feng2018review,
  title={Review of fully coherent free-electron lasers},
  author={Feng, Chao and Deng, Hai-Xiao},
  journal={Nuclear Science and Techniques},
  volume={29},
  number={11},
  pages={160},
  year={2018},
  publisher={Springer}
}

@article{bergues2018tabletop,
  title={Tabletop nonlinear optics in the 100-eV spectral region},
  author={Bergues, Boris and Rivas, DE and Weidman, Matthew and Muschet, AA and Helml, Wolfram and Guggenmos, Alexander and Pervak, V and Kleineberg, Ulf and Marcus, G and Kienberger, Reinhard and others},
  journal={Optica},
  volume={5},
  number={3},
  pages={237--242},
  year={2018},
  publisher={Optica Publishing Group}
}

@article{makos2020alpha,
  title={A 10-gigawatt attosecond source for non-linear XUV optics and XUV-pump-XUV-probe studies},
  author={Makos, Ioannis and Orfanos, I and Nayak, Arjun and Peschel, Jasper and Major, Bal{\'a}zs and Liontos, Ioannis and Skantzakis, Emmanuel and Papadakis, N and Kalpouzos, Constantinos and Dumergue, Mathieu and others},
  journal={Scientific Reports},
  volume={10},
  number={1},
  pages={3759},
  year={2020},
  publisher={Nature Publishing Group UK London}
}

@article{heyl2016scale,
  title={Scale-invariant nonlinear optics in gases},
  author={Heyl, Christoph M and Coudert-Alteirac, Helene and Miranda, Miguel and Louisy, Maite and Kov{\'a}cs, Katalin and Tosa, Valer and Balogh, Emeric and Varj{\'u}, Katalin and L’Huillier, Anne and Couairon, Arnaud and others},
  journal={Optica},
  volume={3},
  number={1},
  pages={75--81},
  year={2016},
  publisher={Optica Publishing Group}
}

@article{wall1990titanium,
  title={Titanium sapphire lasers},
  author={Wall, K. and Sanchez, A.},
  journal={The Lincoln laboratory journal},
  volume={3},
  number={3},
  pages={447--462},
  year={1990},
  publisher={Citeseer}
}

@article{mashiko2008double,
  title={Double optical gating of high-order harmonic generation with carrier-envelope phase stabilized lasers},
  author={Mashiko, Hiroki and Gilbertson, Steve and Li, Chengquan and Khan, Sabih D and Shakya, Mahendra M and Moon, Eric and Chang, Zenghu},
  journal={Physical Review Letters},
  volume={100},
  number={10},
  pages={103906},
  year={2008},
  publisher={APS}
}

@article{nagy2021high,
  title={High-energy few-cycle pulses: post-compression techniques},
  author={Nagy, Tamas and Simon, Peter and Veisz, Laszlo},
  journal={Advances in Physics: X},
  volume={6},
  number={1},
  pages={1845795},
  year={2021},
  publisher={Taylor \& Francis}
}

@article{kuhn2017eli,
  title={The \uppercase{ELI-ALPS} facility: the next generation of attosecond sources},
  author={K{\"u}hn, Sergei and Dumergue, Mathieu and Kahaly, Subhendu and Mondal, Sudipta and F{\"u}le, Mikl{\'o}s and Csizmadia, Tam{\'a}s and Farkas, Bal{\'a}zs and Major, Bal{\'a}zs and V{\'a}rallyay, Zolt{\'a}n and Cormier, Eric and others},
  journal={Journal of Physics B: Atomic, Molecular and Optical Physics},
  volume={50},
  number={13},
  pages={132002},
  year={2017},
  publisher={IOP Publishing}
}

@article{midorikawa2022progress,
  title={Progress on table-top isolated attosecond light sources},
  author={Midorikawa, Katsumi},
  journal={Nature Photonics},
  volume={16},
  number={4},
  pages={267--278},
  year={2022},
  publisher={Nature Publishing Group UK London}
}

@article{rivas2017next,
  title={Next generation driver for attosecond and laser-plasma physics},
  author={Rivas, DE and Borot, A and Cardenas, DE and Marcus, Gilad and Gu, Xun and Herrmann, Daniel and Xu, Jia and Tan, J and Kormin, Dmitrii and Ma, G and others},
  journal={Scientific reports},
  volume={7},
  number={1},
  pages={5224},
  year={2017},
  publisher={Nature Publishing Group UK London}
}

@article{lws100,
  title={Waveform-controlled field synthesis of sub-two-cycle pulses at the 100 \uppercase{TW} peak power level},
  author={Laszlo Veisz and Peter Fischer and Sajjad Vardast and Fritz Schnur and Alexander Muschet and Aitor De Andres and Sreehari Kaniyeri and Hang Li and Roushdey Salh and Kárpát Ferencz and Gergely Norbert Nagy and Subhendu Kahaly},
  journal={Nature Photonics},
  volume={19},
  number={1},
  pages={1013–1019},
  year={2025},
  publisher={Nature Publishing Group UK London}
}

@inproceedings{bergues2017towards,
  title={Towards attosecond XUV-pump XUV-probe measurements in the 100-eV region},
  author={Bergues, B and Rivas, DE and Weidman, M and Muschet, AA and Helml, W and Guggenmos, A and Pervak, V and Matyba, Piotr and Kleineberg, U and Marcus, G and others},
  booktitle={2017 Conference on Lasers and Electro-Optics Europe \& European Quantum Electronics Conference (CLEO/Europe-EQEC)},
  pages={1--1},
  year={2017},
  organization={IEEE}
}

@phdthesis{muschet2021non,
  title={Non-linear attosecond physics at 100 eV},
  author={Muschet, Alexander},
  year={2021},
  school={Ume{\aa} University}
}

@article{abel2009isolated,
  title={Isolated attosecond pulses from ionization gating of high-harmonic emission},
  author={Abel, Mark J and Pfeifer, Thomas and Nagel, Phillip M and Boutu, Willem and Bell, M Justine and Steiner, Colby P and Neumark, Daniel M and Leone, Stephen R},
  journal={Chemical Physics},
  volume={366},
  number={1-3},
  pages={9--14},
  year={2009},
  publisher={Elsevier}
}

@article{sola2006controlling,
  title={Controlling attosecond electron dynamics by phase-stabilized polarization gating},
  author={Sola, IJ and M{\'e}vel, E and Elouga, L and Constant, E and Strelkov, V and Poletto, L and Villoresi, P and Benedetti, Enrico and Caumes, J-P and Stagira, Salvatore and others},
  journal={Nature Physics},
  volume={2},
  number={5},
  pages={319--322},
  year={2006},
  publisher={Nature Publishing Group UK London}
}

@article{lozovoy2008direct,
  title={Direct measurement of spectral phase for ultrashort laser pulses},
  author={Lozovoy, Vadim V and Xu, Bingwei and Coello, Yves and Dantus, Marcos},
  journal={Optics Express},
  volume={16},
  number={2},
  pages={592--597},
  year={2008},
  publisher={Optica Publishing Group}
}

@article{loriot2013self,
  title={Self-referenced characterization of femtosecond laser pulses by chirp scan},
  author={Loriot, Vincent and Gitzinger, Gregory and Forget, Nicolas},
  journal={Optics Express},
  volume={21},
  number={21},
  pages={24879--24893},
  year={2013},
  publisher={Optica Publishing Group}
}

@article{escoto2018advanced,
  title={Advanced phase retrieval for dispersion scan: a comparative study},
  author={Escoto, Esmerando and Tajalli, Ayhan and Nagy, Tamas and Steinmeyer, G{\"u}nter},
  journal={JOSA B},
  volume={35},
  number={1},
  pages={8--19},
  year={2018},
  publisher={Optica Publishing Group}
}

@phdthesis{rivas2016generation,
  title={Generation of intense isolated attosecond pulses at 100 eV},
  author={Rivas, Daniel E},
  year={2016},
  school={lmu}
}

@article{ferancesca2014,
author = {F. Calegari  and D. Ayuso  and A. Trabattoni  and L. Belshaw  and S. De Camillis  and S. Anumula  and F. Frassetto  and L. Poletto  and A. Palacios  and P. Decleva  and J. B. Greenwood  and F. Martín  and M. Nisoli },
title = {Ultrafast electron dynamics in phenylalanine initiated by attosecond pulses},
journal = {Science},
volume = {346},
number = {6207},
pages = {336-339},
year = {2014}
}

@article{schultze2010,
author = {M. Schultze  and M. Fieß  and N. Karpowicz  and J. Gagnon  and M. Korbman  and M. Hofstetter  and S. Neppl  and A. L. Cavalieri  and Y. Komninos  and Th. Mercouris  and C. A. Nicolaides  and R. Pazourek  and S. Nagele  and J. Feist  and J. Burgdörfer  and A. M. Azzeer  and R. Ernstorfer  and R. Kienberger  and U. Kleineberg  and E. Goulielmakis  and F. Krausz  and V. S. Yakovlev },
title = {Delay in Photoemission},
journal = {Science},
volume = {328},
number = {5986},
pages = {1658-1662},
year = {2010}
}

@article{Leshchenko2023,
author = {Vyacheslav Leshchenko and Stephen J. Hageman and Coleman Cariker and Gregory Smith and Antoine Camper and Bradford K. Talbert and Pierre Agostini and Luca Argenti and Louis F. DiMauro},
journal = {Optica},
number = {2},
pages = {142--146},
publisher = {Optica Publishing Group},
title = {Kramers--Kronig relation in attosecond transient absorption spectroscopy},
volume = {10},
month = {Feb},
year = {2023}
}

@article{HENKE1993181,
title = {X-Ray Interactions: Photoabsorption, Scattering, Transmission, and Reflection at E = 50-30,000 eV, Z = 1-92},
author = {B.L. Henke and E.M. Gullikson and J.C. Davis},
journal = {Atomic Data and Nuclear Data Tables},
volume = {54},
number = {2},
pages = {181-342},
year = {1993},
issn = {0092-640X},
}

@article{bourassin2013focus,
  title={How to focus an attosecond pulse},
  author={Bourassin-Bouchet, Charles and Mang, Matthias Maximilian and Delmotte, Franck and Chavel, Pierre and De Rossi, S{\'e}bastien},
  journal={Optics Express},
  volume={21},
  number={2},
  pages={2506--2520},
  year={2013},
  publisher={Optica Publishing Group}
}

@article{haelbich1976multilayer,
  title={Multilayer interference mirrors for the XUV range around 100 eV photon energy},
  author={Haelbich, R-P and Kunz, C},
  journal={Optics Communications},
  volume={17},
  number={3},
  pages={287--292},
  year={1976},
  publisher={Elsevier}
}

@article{hofstetter2011lanthanum,
  title={Lanthanum--molybdenum multilayer mirrors for attosecond pulses between 80 and 130 eV},
  author={Hofstetter, Michael and Aquila, A and Schultze, Martin and Guggenmos, A and Yang, S and Gullikson, E and Huth, M and Nickel, B and Gagnon, Justin and Yakovlev, Vladislav S and others},
  journal={New Journal of Physics},
  volume={13},
  number={6},
  pages={063038},
  year={2011},
  publisher={IOP Publishing}
}

@article{schultze2011spatially,
  title={Spatially resolved measurement of ionization yields in the focus of an intense laser pulse},
  author={Schultze, Martin and Bergues, Boris and Schr{\"o}der, Hartmut and Krausz, Ferenc and Kompa, Karl Ludwig},
  journal={New Journal of Physics},
  volume={13},
  number={3},
  pages={033001},
  year={2011},
  publisher={IOP Publishing}
}

@article{tzallas2018time,
  title={Time gated ion microscopy of light-atom interactions},
  author={Tzallas, P and Bergues, Boris and Rompotis, D and Tsatrafyllis, N and Chatziathanassiou, S and Muschet, Alexander and Veisz, L{\'a}szl{\'o} and Schr{\"o}der, Hartmut and Charalambidis, D},
  journal={Journal of Optics},
  volume={20},
  number={2},
  pages={024018},
  year={2018},
  publisher={IOP Publishing}
}

@article{kolliopoulos2014single,
  title={Single-shot autocorrelator for extreme-ultraviolet radiation},
  author={Kolliopoulos, G and Tzallas, P and Bergues, Boris and Carpeggiani, PA and Heissler, Patrick and Schr{\"o}der, Hartmut and Veisz, L{\'a}szl{\'o} and Charalambidis, D and Tsakiris, George D},
  journal={JOSA B},
  volume={31},
  number={5},
  pages={926--938},
  year={2014},
  publisher={Optica Publishing Group}
}

@article{tsatrafyllis2016ion,
  title={The ion microscope as a tool for quantitative measurements in the extreme ultraviolet},
  author={Tsatrafyllis, N and Bergues, Boris and Schr{\"o}der, Hartmut and Veisz, L{\'a}szl{\'o} and Skantzakis, E and Gray, D and Bodi, B and Kuhn, S and Tsakiris, George D and Charalambidis, D and others},
  journal={Scientific Reports},
  volume={6},
  number={1},
  pages={21556},
  year={2016},
  publisher={Nature Publishing Group UK London}
}

@article{temporalsupperresolution,
  title={Utilizing the temporal superresolution approach in an optical parametric synthesizer to generate multi-TW sub-4-fs light pulses},
  author={A. A. Muschet and A. De Andres and P. Fischer and R. Salh and and L. Veisz},
  journal={Optics Express},
  volume={30},
  number={3},
  pages={4374-4380},
  year={2022},
  publisher={Optica Publishing Group}
}

@article{Eppink1997,
    author = {Eppink, André T. J. B. and Parker, David H.},
    title = {Velocity map imaging of ions and electrons using electrostatic lenses: Application in photoelectron and photofragment ion imaging of molecular oxygen},
    journal = {Review of Scientific Instruments},
    volume = {68},
    number = {9},
    pages = {3477-3484},
    year = {1997},
    month = {09},
    issn = {0034-6748}
}

@ARTICLE{Driver2024,
   author = {Driver, T. and Mountney, M. and Wang, J. and Ortmann, L. and Al-Haddad, A. and Berrah, N. and Bostedt, C. and Champenois, E. G. and DiMauro, L. F. and Duris, J. and Garratt, D. and Glownia, J. M. and Guo, Z. and Haxton, D. and Isele, E. and Ivanov, I. and Ji, J. and Kamalov, A. and Li, S. and Lin, M.-F. and Marangos, J. P. and Obaid, R. and O’Neal, J. T. and Rosenberger, P. and Shivaram, N. H. and Wang, A. L. and Walter, P. and Wolf, T. J. A. and Wörner, H. J. and Zhang, Z. and Bucksbaum, P. H. and Kling, M. F. and Landsman, A. S. and Lucchese, R. R. and Emmanouilidou, A. and Marinelli, A. and Cryan, J. P.},
   title = {Attosecond delays in X-ray molecular ionization},
   journal = {Nature},
   volume = {632},
   pages = {762–767},
   year = {2024}
}

@ARTICLE{Fabris2015,
   author = {Fabris, D. and Witting, T. and Okell, W. and Walke, D. J. and Matia-Hernando, P. and Henkel, J. and Barillot, T. R. and Lein, M. and Marangos, J. P. and Tisch, J. W. G.},
   title = {Synchronized pulses generated at 20~eV and 90~eV for attosecond pump–probe experiments},
   journal = {Nature Photonics},
   volume = {9},
   pages = {383–387},
   year = {2015}
}

@ARTICLE{Takahashi2013,
   author = {Takahashi, E. and Lan, P. and Mücke, O. and Nabekawa, Y. and Midorikawa, K.},
   title = {Attosecond nonlinear optics using gigawatt-scale isolated attosecond pulses},
   journal = {Nat. Commun.},
   volume = {4},
   pages = {2691},
   year = {2013}
}

@ARTICLE{Rudawski2013,
   author = {Rudawski, P. and Heyl, C. M. and Brizuela, F. and Schwenke, J. and Persson, A. and Mansten, E. and Rakowski, R. and Rading, L. and Campi, F. and Kim, B. and Johnsson, P. and L'Huillier, A.},
   title = {A high-flux high-order harmonic source},
   journal = {Rev. Sci. Instrum.},
   volume = {84},
   pages = {073103},
   year = {2013}
}

@article{LHuillier1992,
  title = {Calculations of high-order harmonic-generation processes in xenon at 1064 nm},
  author = {L'Huillier, Anne and Balcou, Philippe and Candel, Sebastien and Schafer, Kenneth J. and Kulander, Kenneth C.},
  journal = {Phys. Rev. A},
  volume = {46},
  issue = {5},
  pages = {2778--2790},
  numpages = {0},
  year = {1992},
  month = {Sep},
  publisher = {American Physical Society}
}

@article{Weissenbilder2022,
title={How to optimize high-order harmonic generation in gases}, volume={4}, number={11}, journal={Nature Reviews Physics}, author={Weissenbilder, R. and Carlström, S. and Rego, L. and Guo, C. and Heyl, C. M. and Smorenburg, P. and Constant, E. and Arnold, C. L. and L’Huillier, A.}, year={2022}, month={Oct}, pages={713–722}}

@article{Lamas2025,
author = {Fernando Ardana-Lamas  and Seth Lucien Cousin  and Juliette Lignieres  and Jens Biegert },
title = {Brilliant Source of 19.2-Attosecond Soft X-ray Pulses below the Atomic Unit of Time},
journal = {Ultrafast Science},
volume = {5},
number = {},
pages = {0128},
year = {2025}
}

@Article{Rading2018,
AUTHOR = {Rading, Linnea and Lahl, Jan and Maclot, Sylvain and Campi, Filippo and Coudert-Alteirac, Hélène and Oostenrijk, Bart and Peschel, Jasper and Wikmark, Hampus and Rudawski, Piotr and Gisselbrecht, Mathieu and Johnsson, Per},
TITLE = {A Versatile Velocity Map Ion-Electron Covariance Imaging Spectrometer for High-Intensity XUV Experiments},
JOURNAL = {Applied Sciences},
VOLUME = {8},
YEAR = {2018},
NUMBER = {6},
ARTICLE-NUMBER = {998},
ISSN = {2076-3417}
}

@article{Manschwetus2016,
  title = {Two-photon double ionization of neon using an intense attosecond pulse train},
  author = {Manschwetus, B. and Rading, L. and Campi, F. and Maclot, S. and Coudert-Alteirac, H. and Lahl, J. and Wikmark, H. and Rudawski, P. and Heyl, C. M. and Farkas, B. and Mohamed, T. and L'Huillier, A. and Johnsson, P.},
  journal = {Phys. Rev. A},
  volume = {93},
  issue = {6},
  pages = {061402},
  numpages = {5},
  year = {2016},
  month = {Jun},
  publisher = {American Physical Society}
}

@article{Squibb2026,
author = {Richard J. Squibb and Massume Zaki and H{\'e}l{\`e}ne Coudert-Alteirac and Sylvain Maclot and Andreas Hult Roos and Nihar Ranjan Behera and Veronica Ideb{\"o}hn and Emelie Olsson and Mulham Al Hashemi and Lukas Antonsson and Antoine Gerbandier and Zehra Bet{\"u}l Duman and Cord Arnold and Per Eng-Johnsson and Anne L'Huillier and Raimund Feifel},
title = {{Attohallen: a new attosecond science facility in Sweden}},
volume = {13537},
booktitle = {Compact Radiation Sources from EUV to Gamma-rays: Development and Applications II},
editor = {Carmen S. Menoni and Jaroslav Nejdl},
organization = {International Society for Optics and Photonics},
journal = {SPIE},
pages = {1353705},
year = {2025}
}

@article{Campi2016,
    author = {Campi, F. and Coudert-Alteirac, H. and Miranda, M. and Rading, L. and Manschwetus, B. and Rudawski, P. and L’Huillier, A. and Johnsson, P.},
    title = {Design and test of a broadband split-and-delay unit for attosecond XUV-XUV pump-probe experiments},
    journal = {Review of Scientific Instruments},
    volume = {87},
    number = {2},
    pages = {023106},
    year = {2016},
    month = {02},
    issn = {0034-6748}
}

@article{Xu2024,
  author    = {L. Xu and E. J. Takahashi},
  title     = {Dual-chirped optical parametric amplification of high-energy single-cycle laser pulses},
  journal   = {Nature Photonics},
  volume    = {18},
  number    = {1},
  pages     = {99--106},
  year      = {2024}
}

@article{Rossi2020,
    title = {{Sub-cycle millijoule-level parametric waveform synthesizer for attosecond science}},
    year = {2020},
    journal = {Nature Photonics},
    author = {Rossi, Giulio Maria and Mainz, Roland E. and Yang, Yudong and Scheiba, Fabian and Silva-Toledo, Miguel A. and Chia, Shih Hsuan and Keathley, Phillip D. and Fang, Shaobo and M{\"{u}}cke, Oliver D. and Manzoni, Cristian and Cerullo, Giulio and Cirmi, Giovanni and K{\"{a}}rtner, Franz X.},
    number = {10},
    pages = {629--635},
    volume = {14},
    publisher = {Springer US},
    issn = {17494893}
}

@article{Palacios2014,
author = {Palacios, Alicia and González-Castrillo, Alberto and  Martín, Fernando},
title = {Molecular interferometer to decode attosecond electron–nuclear dynamics},
journal = {Proceedings of the National Academy of Sciences},
volume = {111},
number = {11},
pages = {3973-3978},
year = {2014}
}

@article{Palacios2009,
author = {Palacios, A. and Rescigno, T. N. and McCurdy, C. W.},
title = {Two-Electron Time-Delay Interference in Atomic Double Ionization by Attosecond Pulses},
journal = {Phys. Rev. Lett.},
volume = {103},
pages = {253001},
year = {2009}
}

@Article{Muschet2022,
AUTHOR = {Muschet, Alexander A. and De Andres, Aitor and Smijesh, N. and Veisz, Laszlo},
TITLE = {An Easy Technique for Focus Characterization and Optimization of XUV and Soft X-ray Pulses},
JOURNAL = {Applied Sciences},
VOLUME = {12},
YEAR = {2022},
NUMBER = {11},
ARTICLE-NUMBER = {5652},
ISSN = {2076-3417}
}

@article{Osolodkov2020,
year = {2020},
month = {sep},
publisher = {IOP Publishing},
volume = {53},
number = {19},
pages = {194003},
author = {Osolodkov, Mikhail and Furch, Federico J and Schell, Felix and Šušnjar, Peter and Cavalcante, Fabio and Menoni, Carmen S and Schulz, Claus P and Witting, Tobias and Vrakking, Marc J J},
title = {Generation and characterisation of few-pulse attosecond pulse trains at 100 kHz repetition rate},
journal = {Journal of Physics B: Atomic, Molecular and Optical Physics}
}

@article{Lorek20214,
author = {Lorek, E. and Larsen, E. W. and Heyl, C. M. and Carlstr{\"o}m, S. and Pale{\v{c}}ek, D. and Zigmantas, D. and Mauritsson, J.},
  title = {High-order harmonic generation using a high-repetition-rate turnkey laser},
  journal = {Review of Scientific Instruments},
  volume = {85},
  number = {12},
  pages = {123106},
  year = {2014}
}

@article{Rothhardt2016,
author = {Jan Rothhardt and Steffen H\"{a}drich and Yariv Shamir and Maxim Tschnernajew and Robert Klas and Armin Hoffmann and Getnet K. Tadesse and Arno Klenke and Thomas Gottschall and Tino Eidam and Jens Limpert and Andreas T\"{u}nnermann and Rebecca Boll and Cedric Bomme and Hatem Dachraoui and Benjamin Erk and Michele Di Fraia and Daniel A. Horke and Thomas Kierspel and Terence Mullins and Andreas Przystawik and Evgeny Savelyev and Joss Wiese and Tim Laarmann and Jochen K\"{u}pper and Daniel Rolles},
journal = {Opt. Express},
number = {16},
pages = {18133--18147},
publisher = {Optica Publishing Group},
title = {High-repetition-rate and high-photon-flux 70 eV high-harmonic source for coincidence ion imaging of gas-phase molecules},
volume = {24},
month = {Aug},
year = {2016}
}

@article{Kretschmar2020,
author = {Martin Kretschmar and Johannes Tuemmler and Bernd Sch\"{u}tte and Andreas Hoffmann and Bj\"{o}rn Senfftleben and Mark Mero and Mario Sauppe and Daniela Rupp and Marc J. J. Vrakking and Ingo Will and Tamas Nagy},
journal = {Opt. Express},
number = {23},
pages = {34574--34585},
publisher = {Optica Publishing Group},
title = {Thin-disk laser-pumped OPCPA system delivering 4.4 TW few-cycle pulses},
volume = {28},
month = {Nov},
year = {2020}
}

@article{Herrmann2009,
author = {Daniel Herrmann and Laszlo Veisz and Raphael Tautz and Franz Tavella and Karl Schmid and Vladimir Pervak and Ferenc Krausz},
journal = {Opt. Lett.},
number = {16},
pages = {2459--2461},
publisher = {Optica Publishing Group},
title = {Generation of sub-three-cycle, 16 TW light pulses by using noncollinear optical parametric chirped-pulse amplification},
volume = {34},
month = {Aug},
year = {2009}
}

@article{Orfanos2019,
  author  = {Orfanos, I. and Makos, I. and Liontos, I. and Skantzakis, E. and F{\"o}rg, B. and Charalambidis, D. and Tzallas, P.},
  title   = {Attosecond pulse metrology},
  journal = {APL Photonics},
  volume  = {4},
  number  = {8},
  pages   = {080901},
  year    = {2019},
  month   = {August}
}

@article{dahl2000simion,
  title={SIMION 3D Version 7.0 User’s manual},
  author={Dahl, David A},
  journal={Idaho National Engineering and Environmental Laboratory, Idaho Falls, ID},
  pages={2--1},
  year={2000}
}

@article{Carstens2016,
author = {H. Carstens and M. H\"{o}gner and T. Saule and S. Holzberger and N. Lilienfein and A. Guggenmos and C. Jocher and T. Eidam and D. Esser and V. Tosa and V. Pervak and J. Limpert and A. T\"{u}nnermann and U. Kleineberg and F. Krausz and I. Pupeza},
journal = {Optica},
number = {4},
pages = {366--369},
publisher = {Optica Publishing Group},
title = {High-harmonic generation at 250 MHz with photon energies exceeding 100 eV},
volume = {3},
month = {Apr},
year = {2016},
}

@ARTICLE{Pupeza2013,
  title     = "Compact high-repetition-rate source of coherent 100 eV radiation",
  author    = "Pupeza, I and Holzberger, S and Eidam, T and Carstens, H and
               Esser, D and Weitenberg, J and Ru{\ss}b{\"u}ldt, P and
               Rauschenberger, J and Limpert, J and Udem, Th and
               T{\"u}nnermann, A and H{\"a}nsch, T W and Apolonski, A and
               Krausz, F and Fill, E",
  journal   = "Nat. Photonics",
  publisher = "Springer Science and Business Media LLC",
  volume    =  7,
  number    =  8,
  pages     = "608--612",
  month     =  aug,
  year      =  2013,
  language  = "en"
}

@article{Li2019,
author = {Jie Li and Andrew Chew and Shuyuan Hu and Jonathon White and Xiaoming Ren and Seunghwoi Han and Yanchun Yin and Yang Wang and Yi Wu and Zenghu Chang},
journal = {Opt. Express},
number = {21},
pages = {30280--30286},
publisher = {Optica Publishing Group},
title = {Double optical gating for generating high flux isolated attosecond pulses in the soft X-ray regime},
volume = {27},
month = {Oct},
year = {2019},
}

@article{Cousin2017,
  title = {Attosecond Streaking in the Water Window: A New Regime of Attosecond Pulse Characterization},
  author = {Cousin, Seth L. and Di Palo, Nicola and Buades, B\'arbara and Teichmann, Stephan M. and Reduzzi, M. and Devetta, M. and Kheifets, A. and Sansone, G. and Biegert, Jens},
  journal = {Phys. Rev. X},
  volume = {7},
  issue = {4},
  pages = {041030},
  numpages = {14},
  year = {2017},
  month = {Nov},
  publisher = {American Physical Society},
}

@article{Timmers2016,
author = {Henry Timmers and Mazyar Sabbar and Johannes Hellwagner and Yuki Kobayashi and Daniel M. Neumark and Stephen R. Leone},
journal = {Optica},
number = {7},
pages = {707--710},
publisher = {Optica Publishing Group},
title = {Polarization-assisted amplitude gating as a route to tunable, high-contrast attosecond pulses},
volume = {3},
month = {Jul},
year = {2016},
}

@article{Mikaelsson2021,
title = {A high-repetition rate attosecond light source for time-resolved coincidence spectroscopy},
author = {Sara Mikaelsson and Jan Vogelsang and Chen Guo and Ivan Sytcevich and Anne-Lise Viotti and Fabian Langer and Yu-Chen Cheng and Saikat Nandi and Wenjie Jin and Anna Olofsson and Robin Weissenbilder and Johan Mauritsson and Anne L’Huillier and Mathieu Gisselbrecht and Cord L. Arnold},
pages = {117--128},
volume = {10},
number = {1},
journal = {Nanophotonics},
year = {2021},
}

@ARTICLE{Harth2018,
  title     = "Compact 200 kHz HHG source driven by a few-cycle OPCPA",
  author    = "Harth, Anne and Guo, Chen and Cheng, Yu-Chen and Losquin, Arthur
               and Miranda, Miguel and Mikaelsson, Sara and Heyl, Christoph M
               and Prochnow, Oliver and Ahrens, Jan and Morgner, Uwe and
               L'Huillier, Anne and Arnold, Cord L",
  journal   = "J. Opt.",
  publisher = "IOP Publishing",
  volume    =  20,
  number    =  1,
  pages     = "014007",
  month     =  jan,
  year      =  2018,
}

@article{Chevreuil2021,
author = {P.-A. Chevreuil and F. Brunner and S. Hrisafov and J. Pupeikis and C. R. Phillips and U. Keller and L. Gallmann},
journal = {Opt. Express},
number = {21},
pages = {32996--33008},
publisher = {Optica Publishing Group},
title = {Water-window high harmonic generation with 0.8-{\textmu}m and 2.2-{\textmu}m OPCPAs at 100 kHz},
volume = {29},
month = {Oct},
year = {2021},
}

@article{Goulielmakis2008,
author = {E. Goulielmakis  and M. Schultze  and M. Hofstetter  and V. S. Yakovlev  and J. Gagnon  and M. Uiberacker  and A. L. Aquila  and E. M. Gullikson  and D. T. Attwood  and R. Kienberger  and F. Krausz  and U. Kleineberg },
title = {Single-Cycle Nonlinear Optics},
journal = {Science},
volume = {320},
number = {5883},
pages = {1614-1617},
year = {2008},
}

@article{Mashiko2008,
  title = {Optimizing the photon flux of double optical gated high-order harmonic spectra},
  author = {Mashiko, Hiroki and Gilbertson, Steve and Li, Chengquan and Moon, Eric and Chang, Zenghu},
  journal = {Phys. Rev. A},
  volume = {77},
  issue = {6},
  pages = {063423},
  numpages = {5},
  year = {2008},
  month = {Jun},
  publisher = {American Physical Society},
}

@article{Johnson2018,
author = {Allan S. Johnson  and Dane R. Austin  and David A. Wood  and Christian Brahms  and Andrew Gregory  and Konstantin B. Holzner  and Sebastian Jarosch  and Esben W. Larsen  and Susan Parker  and Christian S. Strüber  and Peng Ye  and John W. G. Tisch  and Jon P. Marangos },
title = {High-flux soft x-ray harmonic generation from ionization-shaped few-cycle laser pulses},
journal = {Science Advances},
volume = {4},
number = {5},
pages = {eaar3761},
year = {2018},
}

@ARTICLE{Takahashi2004,
  title     = "Low-divergence coherent soft x-ray source at 13 nm by high-order harmonics",
  author    = "Takahashi, Eiji J and Nabekawa, Yasuo and Midorikawa, Katsumi",
  journal   = "Appl. Phys. Lett.",
  publisher = "AIP Publishing",
  volume    =  84,
  number    =  1,
  pages     = "4--6",
  month     =  jan,
  year      =  2004,
}

@ARTICLE{Fu2020,
  title     = "High efficiency ultrafast water-window harmonic generation for single-shot soft X-ray spectroscopy",
  author    = "Fu, Yuxi and Nishimura, Kotaro and Shao, Renzhi and Suda, Akira and Midorikawa, Katsumi and Lan, Pengfei and Takahashi, Eiji J",
  journal   = "Commun. Phys.",
  publisher = "Springer Science and Business Media LLC",
  volume    =  3,
  number    =  1,
  month     =  may,
  year      =  2020,
}

@article{Rivas2018,
author = {D. E. Rivas and B. Major and M. Weidman and W. Helml and G. Marcus and R. Kienberger and D. Charalambidis and P. Tzallas and E. Balogh and K. Kov\'{a}cs and V. Tosa and B. Bergues and K. Varj\'{u} and L. Veisz},
journal = {Optica},
number = {10},
pages = {1283--1289},
publisher = {Optica Publishing Group},
title = {Propagation-enhanced generation of intense high-harmonic continua in the 100-eV spectral region},
volume = {5},
month = {Oct},
year = {2018},
}

\end{document}